\documentclass[traditabstract]{aa} 
\usepackage{calc}
\usepackage{graphicx}
\usepackage{lscape}
\usepackage{natbib}
\bibpunct{(}{)}{;}{a}{}{,} 
\usepackage{hyperref} 
\usepackage{breakurl}
\usepackage{url}
\usepackage{bigstrut} 
\usepackage{txfonts}
\begin{document}

\title{Large-scale environments of binary AGB stars probed by \textit{Herschel}\thanks{{\it Herschel} is an ESA space observatory with science instruments provided by European-led Principal Investigator consortia and with important participation from NASA.}}
\subtitle{I. Morphology statistics and case studies of R~Aquarii and W~Aquilae}

   \author{A.~Mayer
          \inst{1}
          \and A.~Jorissen
          \inst{2}
          \and F.~Kerschbaum
          \inst{1}
          \and R.~Ottensamer
          \inst{1,6}
		  \and W.~Nowotny
          \inst{1}    
          \and N.L.J.~Cox
          \inst{3}   
          \and B.~Aringer
          \inst{1}
          \and J.A.D.L.~Blommaert
          \inst{3,7}
          \and L.~Decin
          \inst{3}
          \and S.~van~Eck
          \inst{2} 
          \and H.-P.~Gail
          \inst{4}
          \and M.A.T.~Groenewegen
          \inst{5}
          \and K.~Kornfeld
          \inst{1} 
          \and M.~Mecina
          \inst{1}
          \and Th.~Posch
          \inst{1}
          \and B.~Vandenbussche
          \inst{3}
          \and  C.~Waelkens
          \inst{3}
          }

\institute{University of Vienna, Department of Astrophysics, T\"urkenschanzstra\ss e 17, 1180 Wien, Austria \\
              \email{a.mayer@univie.ac.at}
\and
Institut d'Astronomie et d'Astrophysique, Universit\'e Libre de Bruxelles, CP. 226, Boulevard du Triomphe, 1050 Brussels, Belgium
\and
Instituut voor Sterrenkunde, KU Leuven, Celestijnenlaan, 200D, 3001 Leuven, Belgium  
\and
Institut f\"ur Theoretische Astrophysik, Zentrum f\"ur Astronomie, Universit\"at Heidelberg, Albert-\"Uberle-Str. 2, 69120 Heidelberg, Germany
\and
Koninklijke Sterrenwacht van Belgi\"e, Ringlaan 3, 1180 Brussels, Belgium
\and
Institute of Computer Vision and Graphics, TU Graz, Infeldgasse 16/II, 8010 Graz, Austria  
\and
Department of Physics and Astrophysics, Vrije Universiteit Brussel, Pleinlaan 2, 1050 Brussels, Belgium   
 }

   \date{Received 21 March 2012 / Accepted 12 November 2012}

  \abstract
   {The \textit{Mass loss of Evolved StarS} (MESS) sample offers a selection of 78 asymptotic giant branch (AGB) stars and red supergiants (RSGs) observed with the PACS photometer on-board Herschel at 70\,$\mu$m and 160\,$\mu$m. For most of these objects, the dusty AGB wind is not spherically symmetric and the wind shape can be subdivided into four classes. In the present paper we concentrate on the influence of a companion on the morphology of the stellar wind. Literature was searched to find binaries in the MESS sample, which are subsequently linked to their wind-morphology class to assert that the binaries are not distributed equally among the classes. 
\\
In the second part of the paper we concentrate on the circumstellar environment of the two prominent objects R~Aqr and \object{W~Aql}. Each shows a characteristic signature of a companion interaction with the stellar wind. For the symbiotic star \object{R~Aqr}, PACS revealed two perfectly opposing arms that in part reflect the previously observed ring-shaped nebula in the optical. However, from the far-IR there is evidence that the emitting region is  elliptical rather than circular. The outline of the wind of W~Aql seems to follow a large Archimedean spiral formed by the orbit of the companion but also shows strong indications of an interaction with the interstellar medium. We investigated the nature of the companion of W~Aql and found that the magnitude of the orbital period supports the size of the spiral outline.} 


\titlerunning{Large-scale environments of binary AGB stars probed by \textit{Herschel}. I.}
\maketitle
%

\section{Introduction}
\label{Introduction}

When low- to intermediate-mass stars evolve off the main sequence and finally up the asymptotic giant branch (AGB) in the Hertzsprung-Russell diagram, they successively strip off their envelope and lose most of their mass in the form of a dust-enriched stellar wind. These dusty outflows are eventually dissipated into the interstellar medium (ISM) where they contribute essentially to the chemical enrichment of our Galaxy since the vast majority of stars evolve through this stage.

In the simplest manner the dusty winds expand with a velocity of 5--15\,km\,s$^{-1}$ as spherically symmetric wind bubbles that push away the surrounding ISM and lead to extended circumstellar envelopes (CSEs) around AGB stars \citep{Young1993b}. However, observations of post-AGB objects in the infrared (IR) and optical \citep{Meixner1999,Ueta2000} showed that at least at the end of the AGB phase the envelopes differ from spherical symmetry while \citet{Olofsson1988} with CO observations and \citet{Kerschbaum2010} in the far-IR found shells that are detached from the star. In addition, asymmetric shells were observed at many wavelengths: UV \citep[e.g.][]{Martin2007,Sahai2010}, 21\,cm \ion{H}{i} \citep[e.g.][]{Gardan2006,Matthews2008,Matthews2011}, CO \citep[e.g.][]{Kahane1996,Castro2010}, and mainly IR \citep[e.g.][]{Ueta2006,Ueta2008,Geise2010,Ladjal2010,Jorissen2011,Cox2012}. 

One major trigger for asymmetries is the presence of a companion to the mass-losing AGB star. The impact of binarity on the shaping of the AGB mass loss depends on the binary separation. In close binaries, where the AGB star fills its Roche lobe \citep{Paczynski1971}, mass transfer is usually considered to be dynamically unstable \citep[e.g.,][]{Ritter1996,Soberman1997,Jorissen2004}, provided that the AGB star is more massive than its companion, which results in the formation of a common envelope when the accreting star cannot adjust fast enough to accommodate the arriving matter \citep[e.g.][]{Livio1988,Taam1989,Soker1991,Hjellming1991,Rasio1996}.  
This common envelope is eventually ejected and the system then looks very much like a planetary nebula 
(most likely bipolar) with a binary (or even coalesced-binary) nucleus \citep[e.g.,][and references therein]{Miszalski2012}. 

A less violent variant of this scenario involves a wider system where it is not the stellar 
photosphere, but rather a dense AGB wind that is filling the Roche lobe.  
\citet{Mohamed2007,Mohamed2011} dubbed this situation ``wind Roche-lobe overflow".
The mass transfer is non conservative and matter escapes from the system through an Archimedean spiral, 
a situation already encountered in hydrodynamical simulations of mass transfer in detached systems by e.g. \citet{Sawada1986}, \citet{Theuns1993} and \citet{Mastrodemos1998,Mastrodemos1999}, where this feature was called ``spiral arm" \citep[see also][]{Huggins2009,Kim2012c,Kim2012,Kim2012b,Wang2012}. Observationally, such large-scale spirals were observed around LL~Peg \citep[AFGL~3068;][]{Mauron2006,Morris2006}, RW~LMi \citep[CIT~6;][]{Dinh2009}, and recently R~Scl \citep{Maercker2012}, and also in a bruised version around \textit{o}~Cet \citep[Mira;][]{Mayer2011}. Estimates of the orbital separation are available for three (AFGL~3068, CIT~6, and Mira) of these four systems, flagging them as wide binaries given their projected separations of 109, 70, and 55\,AU \citep{Morris2006,Dinh2009,Karovska1997}. In the framework of this paper, which interprets Herschel observations, it is important to report the finding by \citet{Edgar2008}. The authors suggested that the spiral shock associated with mass-losing AGB stars in binary systems may anneal the amorphous grains to form crystals, as is often seen in the discs around post-AGB binaries \citep{Gielen2011}. 

Finally, one should mention the frequent occurrence of bipolar jets in binary AGB stars \citep[e.g.,][]{Huggins2007}, such as $o$~Cet \citep{Meaburn2009}, R~Aqr \citep{Wallerstein1980,Kafatos1989}, $\pi^1$~Gru \citep{Sahai1992}, and V~Hya \citep{Hirano2004}.

In this paper, we examine a sub-sample of the 78 AGB stars from the MESS \citep[Mass loss of Evolved StarS;][]{Groenewegen2011} guaranteed-time key program (GTKP) for Herschel and give new observational insights into this topic. In Sect. \ref{Distribution}, we give a short statistical overview of the morphology of circumstellar envelopes of binaries in the sample. The focus is then set on two prominent examples: R~Aqr and W~Aql. Each of them shows remarkable structures in the CSE that point to various forms of interaction. 


\section{Observations and the MESS sample}
\label{Observations}

\subsection{Observations}
The observations presented here were carried out with the Photodetector Array Camera and Spectrometer \citep[PACS;][]{Poglitsch2010} on-board Herschel in 2009--2011. PACS is operating in the far-IR wavelength range in three channels, 70\,$\mu$m (blue channel), 100\,$\mu$m (green channel), and 160\,$\mu$m (red channel) two of which are used simultaneously. Usually, these are the blue and red bands. The pixel size in the two channels is $3\farcs2 \times 3\farcs2$ in the blue band and $6\farcs4 \times 6\farcs4$ in the red band with a typical field of view of $5\arcmin \times 5\arcmin$, but the images presented here are oversampled by a factor of 3.2, resulting in a resolution of 1$\arcsec$ pix$^{-1}$ in the blue and 2$\arcsec$ pix$^{-1}$ in the red band. This image resolution has to be compared to the FWHM of $5\farcs 7$ and $11\farcs 4$ at these wavelengths.

The data processing was performed using the standard reduction steps in the Herschel interactive processing environment (HIPE) and the map-making program \textit{Scanamorphos}. Both are described in detail by \citet{Groenewegen2011} and Roussel (in prep.). In addition, deconvolution was applied to all images. A detailed analysis of this process can be found in \citet{Ottensamer2011}.


\subsection{MESS program}
\label{MESS}

The MESS GTKP \citep{Groenewegen2011} performed with the Herschel Space Observatory \citep{Pilbratt2010} strives to investigate the CSE and the mass loss history of evolved stellar objects, benefits from the high spatial resolution of Herschel's PACS and SPIRE instruments \citep{Poglitsch2010,Griffin2010}. 

Earlier results of this program included the detections of multiple shells together with a bow shock around CW~Leo (IRC~+10216) by \citet{Ladjal2010} and \citet{Decin2011} and of detached shells around AQ~And, U~Ant and TT~Cyg \citep{Kerschbaum2010}. \citet{Jorissen2011} examined the envelopes of TX~Psc and X~Her. The envelopes are massively shaped by the interaction of the stellar winds with the oncoming ISM due to their high space velocity. 

\citet{Mayer2011} discussed the circumstellar environment of \textit{o}~Cet and found it to be shaped by a combination of several processes: wind-ISM interaction, a bipolar outflow, and in the interior, the orbiting motion of the  companion that forms a spiral that appears to be broken. The broken appearance results from the bipolar outflow that pierces the dust shells.


\subsection{MESS sample categories}
\label{MESS_categories}

For this study on binary-induced structures, we concentrate on the 78 AGB stars and RSGs in the MESS sample. The composition of this sample is well-balanced since it contains all major spectral classes of late-type stars. More specifically, the sample consists of 37 C-rich, 32 O-rich stars and 9 S-stars.

Recently, \citet{Cox2012} reviewed the properties of the whole MESS sample with the aim to classify the different CSE morphologies and to relate them to the influence of the interacting ISM. In short, five main morphology classes were identified:

\begin{itemize}
\item the \textit{fermata} type, which is characterised by a strong bow shock in the direction of the space motion of the star; 
\item the \textit{eye} type, or double-fermata, which has two elliptically shaped density-enhanced regions; 
\item the \textit{ring} type, which is in general a detached shell \citep[see e.g.][]{Kerschbaum2010}; 
\item the \textit{irregular} type, which has a completely asymmetric morphology. This can be due, for example, to very strong instabilities that destroy the bow shock or to strong interactions with a companion;
\item a \textit{point source}, which shows no resolved CSE at all.
\end{itemize}
A detailed discussion is provided by \citet{Cox2012} and will not be repeated here. The strong indications of ISM interaction and the wide variety of morphologies motivated us to investigate the possible role of companions that shape the CSE in more detail. A discussion of that role based on the PACS images is the topic of the present paper. We refer to the above-mentioned categories throughout.

\begin{landscape}
\begin{table}\footnotesize
      \caption[]{Properties of the suspected binaries in the MESS sample. The table is subdivided into two parts. The upper part lists the physical binaries, the lower part the visual binaries without indicating physical association. The space motion v$_{\rm LSR}$ is calculated following \citet{Johnson1987}. The adopted solar motion to convert heliocentric motion into LSR motion is $(U,V,W)_{\sun} = (11.10,12.24,7.25)$\,km\,s$^{-1}$ \citep{Schonrich2010}. The objects are ordered by increasing distance. }
      \label{binary_list}
 \centering                                      
      \begin{tabular}{l | l l l l l l l l l}          
\hline\hline      
Object & Morphology & Spec. Type & d & $\dot{M}$ & v$_{\rm w}$ & v$_{\rm LSR}$ & Binary detection & References \bigstrut[t]\\
 & & & [pc] & [10$^{-7}$ $M_{\sun}$\,yr$^{-1}$] & [km\,s$^{-1}$] & [km\,s$^{-1}$] & & \\
 \hline
\object{\textit{o}~Cet} & Fermata \& Irregular & M7\,IIIe & $91 \pm 10$ & 2.5 & 8.1 & 107.7 & visual: $\rho = 0\farcs6$, PA $\approx90\degr$ & (1),(2),(3) \bigstrut[t] \\
\object{$\theta$ Aps} & Fermata & M6.5\,III & $113 \pm 6$ & 1.1 & 4.5 & 34.2 & Hipparcos variability-induced mover (VIM)& (1),(2),(4),(5) \\
\object{EP~Aqr} & Fermata & M8\,IIIv & $114 \pm 8$ & 3.1 & 11.5 & 40.2 & complex CO profile & (1),(2),(6),(7) \\
\object{$\pi^1$~Gru} & Irregular & S6e & $163 \pm 20$ & 8.5 & 30.0 & 11.9 & jets, visual: $\rho = 2\farcs45$, PA $=200.4\degr$ & (1),(2),(8)\\
\object{$o^1$~Ori} & Irregular & M3 & $200 \pm 28$ & $< 0.4$ & - & 38.8 & UV excess & (9),(10),(11) \\
\textbf{R~Aqr} & Irregular & M7\,IIIpev & 214 & 0.6& 16.7 & 39.7 & symbiotic & (12),(13),(14),(15),(16),(17) \\
\object{Y~Lyn} & Point source & M6\,SIb-II & 253 $\pm$ 61 & 5.6 & 5.4 & 13.9 & Hipparcos VIM & (1),(9),(5)\\
\object{TW~Hor} & Point source & C & $322 \pm 38$ & 0.24 & 7.5 & 52.4 & UV excess & (1),(18),(19),(16) \\
\textbf{W~Aql} & Fermata & S6.6 & 340 & 130 & 20.0 & 21.6 & composite spectrum/CO profile,  & (20),(2),(1),(21),(18),(22)\\
 & & & & & & & visual: $\rho = 0\farcs47$, PA $= 209\degr$ & \\
\object{R~Scl} & Fermata \& Ring & C & 370 & 16 & 17.0& 66.3 & spiral arm & (2),(1)(23) \\
\object{VY~UMa} & Eye & C & $380 \pm 51$ & 0.7 & 7.9 & 28.2 & UV excess & (1),(4),(24),(10) \\
\object{TX~Cam} & Point source & M8.5 & 380 & 65 & 21.2 & - & spiral arm & (2),(19)\\
\object{U~Cam} & Eye & C & 430 & 10 & 20.6 & 9.8 & Hipparcos DMSA/C: $\rho = 0\farcs17$, PA $= 79\degr$ & (25),(26),(27),(1)\\
\object{V~Eri} & Point source & M5/M6\,IV & $439 \pm 133$ & 1.5 & 13 & 39.8 & UV excess & (1),(26),(24),(10)\\ 
\object{RW~LMi} & Point source & C & 440 & 59 & 20.8 & - & spiral arm& (6),(2),(28) \\
\object{VY~CMa} & Point source & M3/M4\,III & 562 & 2800 & 46.5 & 42.9 & UV excess & (26),(2),(1),(10) \\
\object{V~Hya} & Point source & C & 600 & 610 & 30 & 27.9 & rapid rotation & (14),(2),(1),(29) \\
\object{LL~Peg} & Point source & C & 980 & 310 & 16.0 & - & spiral arm, visual: $\rho = 0\farcs11$ & (2),(30),(31) \bigstrut[b] \\
\hline
\object{R~Dor} & Point source & M8\,IIIe & 55 $\pm$ 3 & 6.1 & 6.0 & 39.4 & visual: $\rho = 32\farcs3$ & (32),(33),(34) \bigstrut[t] \\
\object{W~Hya} & Ring & C & 104 $\pm$ 12 & 0.78 & 7.5 & 48.2 & visual: $\rho = 74\farcs5$ & (1),(2),(34) \\
\object{R~Hya} & Fermata & M7\,IIIe & 124 $\pm$ 11 & 1.6 & 12.5 & 25.3 & visual: $\rho = 21\farcs2$ & (1),(2),(34) \\
\object{R~Cas} & Fermata & M7\,IIIe & 127 $\pm$ 16 & 4.0 & 13.5 & 44.0 & visual: $\rho = 27\farcs8$ & (1),(2),(34) \\
\object{R~Crt} & Eye & M7\,III & 261 $\pm$ 65 & 6.9 & 10.8 & 23.8 & visual: $\rho = 65\farcs4$ & (1),(2),(34) \\
\object{VX~Sgr} & Point source & M5/M6\,III & 262 $\pm$ 187 & 610 & 24.3 & 8.5 & visual: $\rho = 0\farcs7$ & (1),(2),(34) \\
\object{UU~Aur} & Fermata & C & 341 & 2.7 & 11.0 & 17.9 & visual: $\rho = 18\farcs2$ & (32),(2),(34) \\
\object{V~Pav} & Eye & C & 370 $\pm$ 73 & 3.4 & 16.0 & 24.0 & visual: $\rho = 15\farcs0$ & (1),(14),(24),(34) \\
\object{S~Sct} & Ring & C & 386 $\pm$ 85 & 1.4 & 8.0 & 17.5 & visual: $\rho = 14\farcs3$ & (1),(2),(34)\\
\object{TT~Cyg} & Ring & C & 436 & 50 & 12.6 & 38.3 & visual: $\rho = 71\farcs0$ & (32),(35),(36),(34) \bigstrut[b] \\
\hline
\end{tabular}

\tablebib{(1):~\citet{vanLeeuwen2007}, (2):~\citet{DeBeck2010}, (3):~\citet{Karovska1997}, (4):~\citet{Schoier2001}, (5):~\citet{Pourbaix2003}, (6):~\citet{Knapp1998}, (7):~\citet{Winters2007}, (8):~\citet{Sahai1992}, (9):~\citet{Groenewegen1998}, (10):~\citet{Sahai2008}, (11):~\citet{Ake1988}, (12):~\citet{Kamohara2010}, (13):~\citet{Matthews2007}, (14):~\citet{Groenewegen2002}, (15):~\citet{Willson1981}, (16):~\citet{Hinkle1989}, (17):~\citet{Gromadzki2009}, (18):~\citet{Culver1974}, (19):~\citet{Castro2010}, (20):~\citet{Guandalini2008}, (21):~\citet{Herbig1965},  (22):~\citet{Ramstedt2011}, (23):~\citet{Maercker2012}, (24):~\citet{Loup1993}, (25):~\citet{Knapp2003}, (26):~\citet{Young1993b}, (27):~\citet{Lindqvist1999}, (28):~\citet{Dinh2009}, (29):~\citet{Barnbaum1995}, (30):~\citet{Mauron2006}, (31):~\citet{Morris2006}, (32):~\citet{Cox2012}, (33):~\citet{Olofsson2002}, (34):~\citet{Proust1981},  (35):~\citet{Olofsson1996}, (36):~\citet{Olofsson2000}}

\tablefoot{Although VX~Sgr is associated with entry ADS~11078 in \citet{Proust1981}, the Washington Double Star Catalog (\url{http://ad.usno.navy.mil/wds/wdstext.html}) does not list any measurement for that system, which is thus most likely spurious.}
\end{table}
\end{landscape}


\section{Physical binaries in the MESS sample}
\label{Binaries_MESS}

As described above, the FWHM of the PSF is $5\farcs 7$ and $11\farcs 4$ at 70\,$\mu$m and 160\,$\mu$m, respectively. However, these resolutions are insufficient to resolve the individual components of a binary system even for relatively nearby objects. For this reason, the literature was thoroughly searched to find other possible evidence for binarity among the 78 AGB stars and RSGs in the MESS sample. The result of this search is listed in Table \ref{binary_list}, resulting in 18 objects that show strong indications for physical binarity in their spectra, proper motion history, or observed morphology. In addition, the table also lists the objects of the MESS sample that are part of the visual binary catalogue by \citet{Proust1981}, but we stress that most of these visual binaries are likely to be optical pairs given their large angular separations.

Table \ref{binary_list} also provides the general and kinematic data of the targets. In the penultimate column, the detection method of the binarity is listed.\\


\subsection{Difficulties and limitations of binary detection}
\label{Binary_detection}
Flagging an AGB star as a binary is a difficult endeavour, mainly because the main binary-detection method -- variable radial velocities -- is not very efficient in the case of AGB stars (no convincing case of a spectroscopic binary involving a genuine AGB star\footnote{a star is defined as an AGB star if it is a Mira variable, carbon-rich, or exhibits spectral lines from the unstable element technetium} 
is currently known). This lack of efficiency results from the fact that orbital radial-velocity variations are confused by the intrinsic
variations associated with the atmospheric dynamics 
\citep[e.g.,][]{Hinkle1982,Hinkle2002,Nowotny2010}, especially since the large radii of AGB stars impose their orbits to be rather wide, and hence the velocity amplitude to be small. 

Other methods of binary detection should thus be used in the case of AGB stars, of which we list:
\begin{itemize}
\item detection of the companion itself (visually, interferometrically or through lunar occultations); 
\item composite nature of the spectrum and/or symbiotic activity; 
\item X-ray emission;
\item shallow light curve of a Mira variable, with wide and flat minima; 
\item light variations associated with orbital motion (eclipses, Algol);
\item asymmetries in circumstellar envelopes (especially spiral features, or double CO line profile);
\item proper-motion variations. 
\end{itemize}

A detailed description of these methods (with appropriate references) may be found in \citet{Jorissen2004} and  \citet{Jorissen2008}. The selection of putative binaries among the MESS targets is based on these different methods, as indicated in the penultimate column of Table~\ref{binary_list}.  
The confidence level for a target to be a possible binary varies from one target to another. In the case of visual binaries with a small angular separation (o~Cet, $\pi^1$~Gru, W~Aql), there can be no doubt; there is very little doubt in the case of  UV-excess detection ($o^1$~Ori, VY~UMa, TW~Hor, V~Eri), but the confidence is lowest for the double CO profile (EP~Aqr). To account for these profiles, a bipolar geometry accompanying a Keplerian circumbinary disc has been proposed for instance for BM~Gem \citep{Kahane1996}, whose companion was later detected by \citet{Izumiura2008}. But a double wind (fast and slow), as originally suggested by \citet{Knapp1998} for instance for the MESS target EP~Aqr is an alternative to the binary model.  The recent CO map of the circumstellar environment of that object by \citet{Nakashima2006} casts doubt on the binary nature of that star, since it does not reveal any signature of a circumbinary disc. In those cases, other binary diagnostics, such as the effect of binarity on the shaping of the CSE can dispel doubts.
\begin{table}[t]\footnotesize
      \caption[]{Distribution of morphologies in the MESS sample.}
      \label{distribution_list}
      \begin{tabular}{l | r r r r r r}          
\hline
\hline      
Type & \multicolumn{2}{c}{Number} & \multicolumn{2}{c}{Number} & \multicolumn{2}{c}{Physical binaries} \bigstrut[t] \\
 & \multicolumn{2}{c}{ } & \multicolumn{2}{c}{within 500\,pc} & \multicolumn{2}{c}{within 500\,pc} \bigstrut[b] \\
\hline
Fermata & 25 & [32.1\,\%] & 23 & [39.0\,\%] & 5/23 & [21.7\,\%] \bigstrut[t] \\
Eye & 7 & [9.0\,\%] & 7 & [11.9\,\%] & 2/7 & [28.6\,\%]   \\
Ring & 15 & [19.2\,\%] & 13 & [22.0\,\%] & 1/13 & [7.7\,\%] \\
Irregular & 8 & [10.3\,\%] & 7 & [11.9\,\%] & 4/7 & [57.1\,\%] \\
Point source & 30 & [38.5\,\%] & 15 & [25.4\,\%] & 5/15 & [33.3\,\%] \bigstrut[b] \\
\hline
Total & 78 & & 59 & & 15/59 & [25.4\,\%] \bigstrut[t] \\
\hline
\end{tabular}
\tablefoot{Seven objects, of which two are binaries (\textit{o}~Cet and R~Scl), show multiple morphologies and are counted for two types. Of these seven objects, six are within 500\,pc. For details see \citet{Cox2012}.}
\end{table}


\subsection{Distribution of binaries according to morphological categories}
\label{Distribution}

In this section, we investigate the statistical frequency of binaries among the different morphological categories to see whether some categories are more often associated with binaries than others. \\
Before embarking on this comparison, we emphasize, however, that our ability to detect complex structures around AGB stars depends upon their distance, since it is obviously easier for nearby objects. This is reflected in the statistics derived from the \citet{Cox2012} morphologies: the number of point sources drops from 30/78 (38.5\,\% of the total sample) to 15/59 (25.4\,\%) for the d$<$500\,pc objects, as seen in Table \ref{distribution_list}. For this reason, the binary frequency among the different morphologies is only investigated for the 59 objects within 500\,pc from the Sun. 

Only three among the 18 binaries are farther away than 500\,pc, which yields a binary frequency of 25.4\,\% in the MESS d$<$500\,pc sample. This value is not so different from the 30\,\% of binaries detected by \citet{vanWinckel2003} among post-AGB stars.

Furthermore, seven objects of the MESS sample show multiple morphologies, and two of them are binary stars: \textit{o}~Cet, whose extended structure was already discussed in detail by \citet{Mayer2011}, and R~Scl \citep{Maercker2012}. 

The most common morphology type within 500\,pc of the sample is the \textit{fermata}: 23 out of the 59 objects (39.0\,\%) show this kind of strong AGB wind -- ISM interaction. Of these 23 objects, 5 (or 21.7\,\%) are physical binaries. Do these five binary stars have a fermata morphology, which would suggest that the main shaping agent is indeed the interaction of the fast moving AGB wind with the ISM, which would mean in turn that the binary motion is negligible? In fact, the classification is not as clear-cut as it might seem. For instance, the binary nature of EP~Aqr is subject to debate, as discussed in Sect.~\ref{Binary_detection}, and should thus not be included in the present discussion. For $o$~Cet and R~Scl, fermata is not the sole morphological type, because \citet{Cox2012} also associate them with types ``irregular" and ``ring", respectively. Since $o$~Cet is the nearest binary target, it is not really surprising that this system shows the most conspicuous impact of the binary motion, in the form of broken arcs and of a symmetry axis tilted with respect to the space motion \citep{Mayer2011}. $\theta$~Aps and W~Aql, like $o$~Cet, are other cases where signatures of binary motion and ISM interaction seem to co-exist in the CSE appearance: we show in Sect.~\ref{W_Aql} that an Archimedean spiral fits the CSE of W~Aql well, even though the spiral also shows some signs of interaction with the ISM. In a forthcoming paper, we will demonstrate that the CSE of $\theta$~Aps shows indications for a jet bumping into the wind -– ISM bow shock. Thus, there is not one single pure fermata morphology among the binary stars.

The \textit{eye} type is a comparatively small group of seven objects (representing 11.9\,\% of the d$<$500\,pc sample) two of which are binaries (28.6\,\%). Given the small-number statistics, no conclusion can be drawn for this category.

The \textit{ring} category contains 13 members within 500\,pc (22.0\,\%). Ring objects in general are slowly moving through the ISM, and most of them are carbon stars. In some cases, however, a bow shock forms beyond the ring structure, supporting the assumption that the rings are rather due to wind-wind interaction, probably associated to the transition from an O-rich to a C-rich atmosphere \citep{Cox2012}. If for one reason or another, the wind from the C-star becomes faster than the wind formerly present around the O-star (for instance because of the change in opacity of the carbon grains with respect to oxygen-based grains, or because of the increase in luminosity accompanying the thermal pulse responsible for the carbon synthesis), the fast wind will catch the old slower wind and interact with it \citep{Mattsson2007}, probably forming the observed ring. Currently, only one (R~Scl) of the 13 ring objects shows indications of binarity, in the form of an Archimedean spiral \citep{Maercker2012}. 

The shaping influence of binarity is best revealed by the \textit{irregular} type stars. Their circumstellar environment exhibits no symmetry and cannot be related to ISM or to wind-wind interaction. There are seven objects belonging to the irregular category, four of which are binaries (57.1\,\%), which is by far the highest percentage. The number of objects is too small though to conclude that binaries are the only cause for an irregular morphology. The four binary objects in the irregular category are $o$~Cet, $o^1$~Ori, $\pi^1$~Gru, and R~Aqr. $o$~Cet was already discussed by \citet{Mayer2011} and for $o^1$~Ori there is doubt whether the very faint emission seen at the sky noise level with PACS is indeed real and associated to the central star \citep[see Fig. 4. in][]{Cox2012}. The CSE of $\pi^1$~Gru shows an arm structure, which is strong evidence for the influence of a companion. This target will be discussed together with $\theta$~Aps in a forthcoming paper.  The IR emission of R~Aqr, dominated by two opposing arms, is discussed in Sect.~\ref{R_Aqr}. The remaining stars in the irregular category with no evidence so far for binarity are AFGL~278, ST~Her, and V~Cyg \citep{Cox2012}. Despite the small number of objects, it is possible to compare the binary frequency in the whole sample ($p_{\rm B} = 15/59 = 0.254$) with that in the ``irregular" category ($p_{\rm B,irr} = 4/7 = 0.571$) and check the null hypothesis $H_0$ that the two frequencies are indeed identical and their difference only due to sampling fluctuations. If the null hypothesis is true, the expected number of binaries among irregular morphologies is
\begin{equation}
\tilde{x} \equiv \tilde{N}_{\rm B,irr} = p_{\rm B}N_{\rm irr} = 1.78,
\end{equation}
as may be estimated from the fraction of binaries in the total sample (denoted $p_{\rm B}$) applied to the number of stars in the irregular category ($N_{\rm irr}$). The corresponding standard deviation on $\tilde{x}$ (denoted $\sigma_x$) is computed from its expression for a hypergeometric distribution, which characterises the variable $\tilde{x}$, given the total number of stars in the sample, $N_{\rm all}$, the total number of binaries, $p_{\rm B}N_{\rm all}$, and the number of stars with irregular morphologies, $N_{\rm irr}$: 
\begin{equation}
\sigma_x = [N_{\rm irr}p_{\rm B}(1-p_{\rm B})(N_{\rm all}-N_{\rm irr})/(N_{\rm all}-1)]^{1/2} = 1.08.
\end{equation}
The reduced, squared difference $c^2$ between the estimated number of binaries with irregular morphologies ($\tilde{x} = 1.78$) and the observed number ($x = 4$) is 
\begin{equation}
c^2 = [(\tilde{x}-x)/\sigma_x]^2 = 4.225,
\end{equation}
and its significance is computed from the $\chi^2$ distribution with one degree of freedom.
The corresponding significance level is then 
\begin{equation}
\alpha = 0.5~P(\chi^2 > c^2) = 0.02.
\end{equation}
Thus, the probability that the difference between the expected number of binaries with irregular morphologies (1.78) and the observed number (4) is due to random fluctuations amounts to only 2\,\%, and may be considered as insignificant.

The last type of objects are the \textit{point sources}, representing 15 of the 59 objects within 500\,pc (25.4\,\%). Five of them are binaries (33.3\,\%), which is about the average percentage of binaries in the whole sample (23.7\,\%).

Finally, we stress that the sample selection, as sketched by \citet{Cox2012}, should not interfere with the above statistics, since that statistics is based on the distribution of binaries among morphological categories defined \textit{a posteriori}.
\\
\\
According to these results, a companion to an AGB star would most likely produce circumstellar structures that are ascribed to the irregular morphology type. However, a closer look at the images, which we report in Sect.~\ref{case_studies} for R~Aqr and W~Aql, leads to the conclusion that the irregular category hides morphologies specific to binary systems, such as jets (R~Aqr) or spiral arms \citep[W~Aql and $o$~Cet;][for the latter]{Mayer2011}. But apparently nothing prevents this binary-induced morphology from coexisting with a bow shock that might result from interaction with the ISM, which is observed on a larger scale, as in the latter two systems.
\begin{figure*}[t!]
\centering
   \includegraphics[width=9cm]{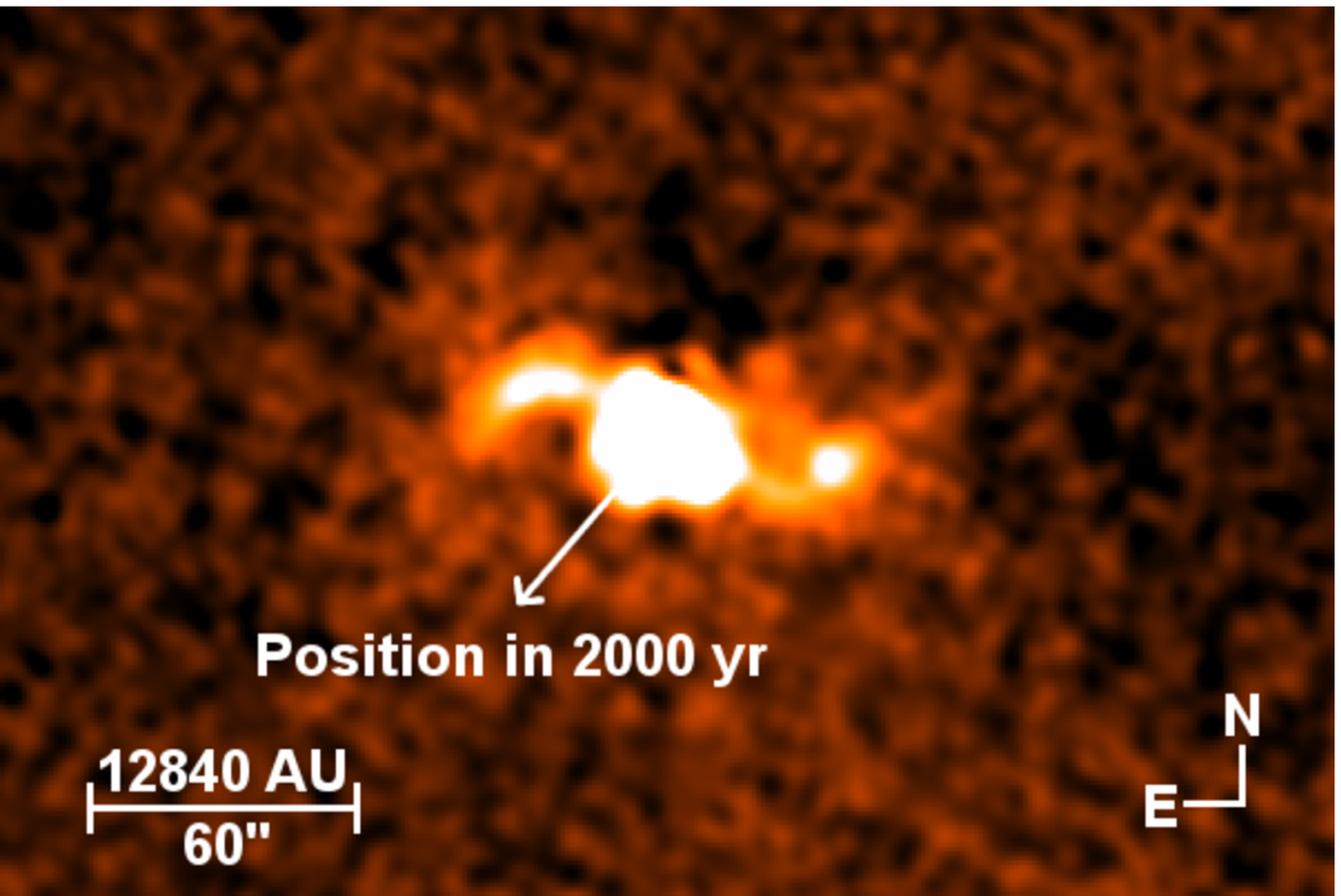} 
   \includegraphics[width=9cm]{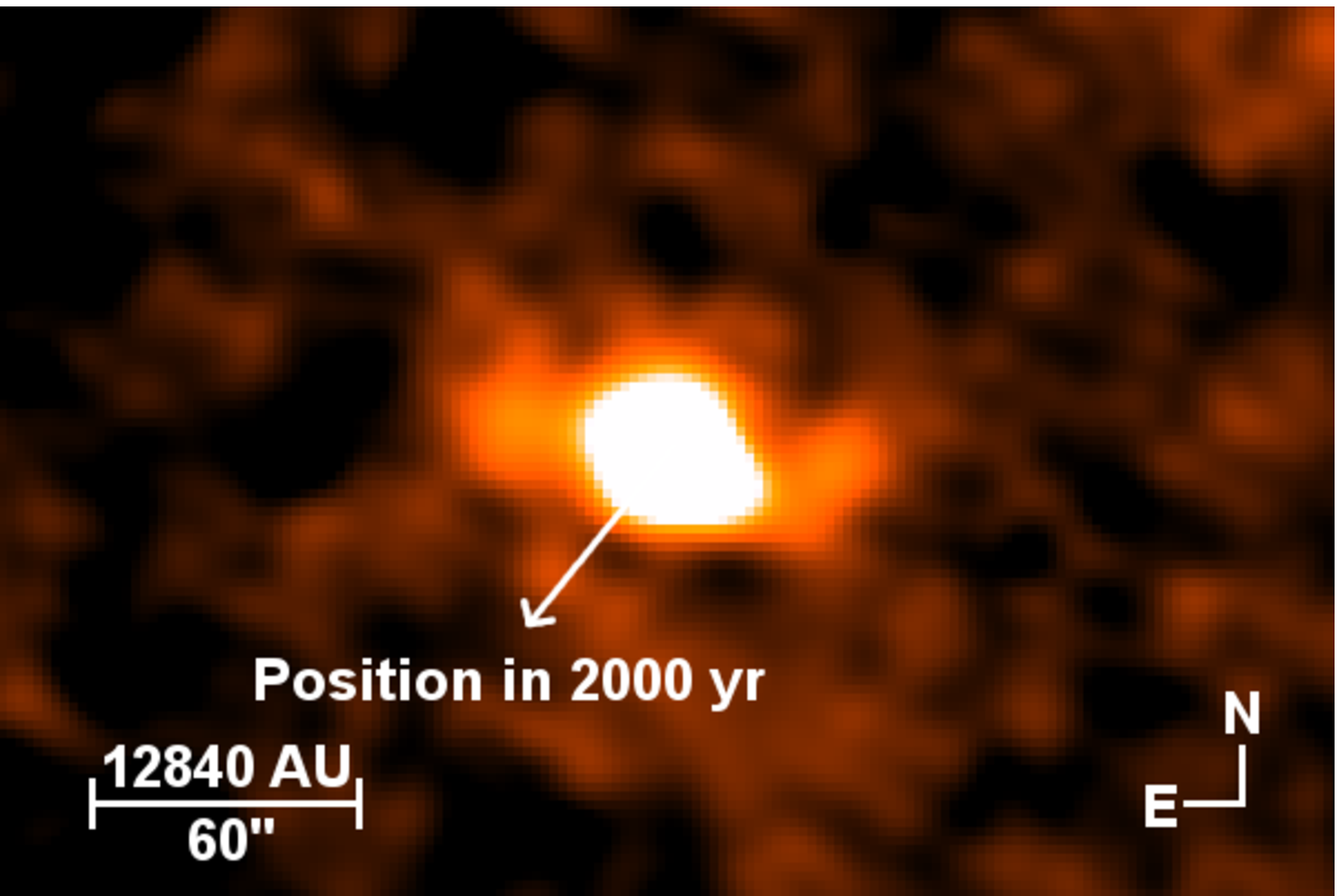}
     \caption{Deconvolved PACS images of R~Aqr at 70\,$\mu$m (left) and at 160\,$\mu$m (right). The arrows indicate the direction of space motion and the position of the star in 2000\,yr. The proper motion values are taken from \citet{Kamohara2010}.}
\label{R_Aqr_deconv}
\end{figure*}


\section{Case studies of binary AGB stellar winds: R~Aqr and W~Aql}
\label{case_studies}

In this section, the new Herschel-PACS far-IR observations of the large-scale environment of R~Aqr and W~Aql are discussed with emphasis on how the binary interplay shapes the AGB wind morphology. These objects were selected because of the complex morphology of their envelope, which suggests a prominent role of binary interaction. This section attempts to elucidate that role.

\subsection{R~Aqr}
\label{R_Aqr}

\subsubsection{Background}
\label{R_Aqr_background}

R~Aqr is one of the best-studied and brightest symbiotic systems \citep{Whitelock1983}, probably observed for almost 1000 years, as suggested by a nova record in Korean history books \citep[1073--1074 A.D.;][]{Yang2005}. The system consists of an M7III Mira variable primary star with a pulsation period of 387 days \citep{Mattei1979} and a hot companion \citep{Merrill1921,Merrill1950} that is most probably a magnetic white dwarf with $M \approx 0.6 - 1$\,$M_{\sun}$ \citep{Nichols2007,Gromadzki2009}. The companion was resolved by \citet{Hollis1997} with a separation of $55\pm 2$~mas from the primary (11.8\,AU at 214\,pc) at a position angle of $18\degr \pm 2\degr$. 

The orbital period of 44\,yr was first determined by \citet{Willson1981} from eclipses in the light curves in 1928--1934 and 1974--1980 and was refined later to 43.6\,yr \citep{Gromadzki2009}. The orientation of the orbit during the eclipse is uncertain, however. From the duration of the eclipse, \citet{Makinen2004} claimed that it occurs at apastron, while \citet{McIntosh2007} suggested that it happens during periastron passage. The latter authors also obtained an orbital period of 34.6\,yr, significantly shorter than the other estimates. 
\begin{table*}[t!]\footnotesize
      \caption[]{Proper motion values of R~Aqr and W~Aql from various catalogues sorted according to their mean epoch. $\mu_{\alpha*}$ stands for $\mu_{\alpha}\, \cos\delta$. }
      \label{proper_motion_table}
      \begin{tabular}{r | r r r r r r r r r}          
\hline\hline   
\textbf{R~Aqr} & GC$^{(1)}$ & SAO$^{(2)}$ & Tycho-2$^{(3)}$ & UCAC3$^{(4)}$ & PPMXL$^{(5)}$ & USNO$^{(6)}$ & HIP2$^{(7)}$ & ASCC-2.5$^{(8)}$  & SiO maser$^{(9)}$ \bigstrut[t] \\
 & 1903.2 & 1903.2 & & 1991.27 & 1994.60 & 2000.0 & & & \\
\hline
$\mu_{\alpha*}$ [mas/yr] & 27 & 29 $\pm$ 5 & 33 $\pm$ 1.4 & 30.8 $\pm$ 1.6 & 28.6 $\pm$ 1.7 & 32 & 33.00 $\pm$ 1.53 & 28.95 $\pm$ 1.52 & 32.2 $\pm$ 0.8\\
$\mu_{\delta}$ [mas/yr] & -20 & -21 $\pm$ 5 & -32.6 $\pm$ 1.2 & -31.5 $\pm$ 2.1 & -31.8 $\pm$ 1.6 & -34 & -25.73 $\pm$ 1.30 & -32.61 $\pm$ 1.85 & -29.5 $\pm$ 0.7 \bigstrut[b] \\
\hline\hline
\textbf{W~Aql} & GC$^{(1)}$ & SAO$^{(2)}$ & YZC$^{(10)}$ & W550$^{(11)}$ & CPIRSS$^{(12)}$ & UCAC3$^{(4)}$ & CMC$^{(13)}$ & ASCC-2.5$^{(7)}$ & PPMXL$^{(6)}$ \bigstrut[t] \\
 & 1913.2 & 1913.2 & 1933.60 & 1968.55 &  & 1981.86 & 1993.68 & & 2002.26 \\
\hline
$\mu_{\alpha*}$ [mas/yr] & 27 & 30 $\pm$ 20 & 26 & 30 & 33 & -5.2 $\pm$ 48.1 & 15.9 $\pm$ 1.8 & 15.06 $\pm$ 1.70 & -21.0 $\pm$ 20.8\\
$\mu_{\delta}$ [mas/yr] & 7 & 8 $\pm$ 22 & 0 & 10 & 8 & 17.6 $\pm$ 48.1 & -0.9 $\pm$ 2.0 & -0.58 $\pm$ 2.00 & - 36.5 $\pm$ 20.8 \bigstrut[b] \\
\hline
\end{tabular}
\tablebib{(1):~General Catalogue of Stars \citep{Boss1937}, (2):~Smithsonian Astrophysical Observatory Star Catalog \citep{SAO1966}, (3):~Tycho-2 \citep{Hog2000}, (4):~Third U.S. Naval Observatory CCD Astrograph Catalog \citep[][based on Hipparcos and Tycho-2 for bright and SPM and SuperCosmos for faint stars]{Zacharias2010}, (5):~PPMXL \citep[][combining USNO-B1.0 and 2MASS]{Roeser2010}, (6):~USNO-B1.0 \citep[][taken from 7435 Schmidt plates over the last 50\,yr]{Monet2003}, (7):~New Hipparcos Reduction \citep{vanLeeuwen2007}, (8):~All-Sky Compiled Catalogue-2.5 \citep[][merging e.g. Hipparcos, Tycho-1, Tycho-2 and PPM]{Kharchenko2009}, (9):~\citet{Kamohara2010}, (10):~Yale Zone Catalogue \citep{Fallon1983}, (11):~Washington 550, Results of observations with the six-inch transit circle 1963-1971 \citep{Hughes1982}, (12):~The U.S. Naval Observatory Catalog of Positions of Infrared Stellar Sources \citep{Hindsley1994}, (13):~Carlsberg Meridian Catalogs Number 1-11 \citep{Carlsberg1999}}
\end{table*}

The orbit is highly eccentric, with $e \approx 0.6-0.8$, and it is seen nearly edge-on with $i = 70\degr$ as noted by \citet{Solf1985}, \citet{Wallerstein1986}, \citet{Hinkle1989}, and \citet{Hollis1997}. 

The distance to the system is fairly well established and ranges from 363\,pc \citep[Hipparcos;][]{vanLeeuwen2007} to 214\,pc \citep[SiO maser spots;][]{Kamohara2010}, with an intermediate value at 240\,pc \citep[period-luminosity relation;][]{Whitelock2008}. Throughout this discussion, the distance of 214\,pc is used since \citet{Kamohara2010} were also able to derive new proper motion values with a small error bar ($\mu_{\alpha} \cos\delta = 32.2 \pm 0.8\,{\rm mas}\,{\rm yr}^{-1}$ and $\mu_{\delta} = -29.5 \pm 0.7\,{\rm mas}\,{\rm yr}^{-1}$). 

In the optical, the system is surrounded by two jets and a nebular structure with the shape of a double hourglass \citep{Solf1985}. The NE optical jet (PA = 46$\degr$) was discovered first \citep{Wallerstein1980} at a distance of about 6$\arcsec$ from the binary system, while  the second (SW) jet was discovered as knots in the radio emission \citep{Kafatos1989}. Subsequently, the jets were intensively studied \citep[e.g.][]{Hollis1990,Hollis1991,Viotti1987,Makinen2004,Nichols2007}. \citet{Kellogg2007} observed both jets in X-ray and radio emission and found the NE jet to be curved and dominated by a bright spot 8$\arcsec$ from the star. The bright spot is tilted away from the observer \citep{Nichols2009}. The SW jet, with PA = 211$\degr$, is not directly opposite, with a bright spot 26$\arcsec$ away from the binary system \citep{Kellogg2007}.

\begin{figure}[t!]
\centering
   \includegraphics[width=9cm]{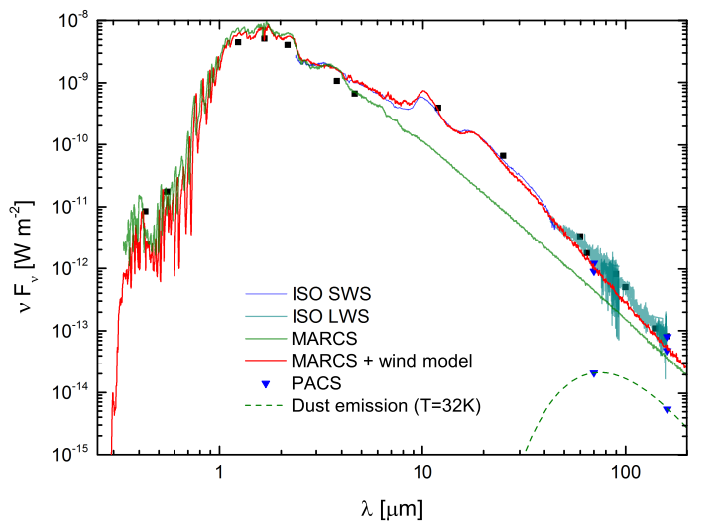} \\
     \caption{SED of R~Aqr described by single-band photometry (filled squares) compiled from the literature as well as the available ISO-SWS (blue line) and -LWS (cyan line) spectra. Overplotted are the synthetic spectra based on models for the photosphere (MARCS; green line) and the combined photosphere + wind (red line; see Section \ref{R_Aqr_spectra} for details). In addition, aperture photometry of the PACS 70\,$\mu$m and 160\,$\mu$m bands (triangles) is included. There are three data points for each of these measurements. From bottom to top: aperture photometry (i) of the eastern arm, (ii) of the central source, and (iii) of the whole IR-emitting region. The dotted line represents the model for the dust emission of the eastern arm.}
\label{R_Aqr_SED}
\end{figure}

Interestingly, \citet{Makinen2004} found that parts of the NE jet broke apart within a timespan of 7\,yr probably due to a collision with a more massive cloud.

In addition to the jet, the circumstellar environment features an hourglass structure \citep[discovered by][]{Solf1985}, extending to about 2$\arcmin$. It is centred on the binary system and might be linked to a possible previous explosion on the white dwarf 660\,yr ago. The structure of the nebula consists of a ring with a radius of 42$\arcsec$ inclined by 18$\degr$ with respect to the plane of the sky. \citet{Hollis1989} compared 6-cm radio continuum emission with observations in the optical and found them to be morphologically similar and spatially related.

\subsubsection{Characterising R~Aqr and its inner wind region}
\label{R_Aqr_spectra}

To understand Herschel's observations of the outer regions of the environment of R~Aqr, it is important to derive characteristics of the star, its present-day mass loss and the region where the outflow is triggered. For this reason, the literature was searched to derive a spectral energy distribution (SED) of R~Aqr. Photometric data were adopted from the ASCC-2.5 catalogue ($B, V$ bands), the 2MASS catalogue ($J, H, K$ bands), \citet{Fouque1992} [$L, M$ bands], the Infrared Astronomical Satellite (IRAS; 12, 25, 60 \& 100\,$\mu$m), and the AKARI satellite (65, 90, 140 \& 160\,$\mu$m). Valuable information is provided by the spectra obtained by the Infrared Space Observatory [ISO; 2--45\,$\mu$m (SWS) and 43--160\,$\mu$m (LWS)]. Figure~\ref{R_Aqr_SED} shows the resulting SED of R~Aqr. The short-wavelength band (SWS) of ISO is dominated by pronounced emission of an optically thin dust shell that contains amorphous silicate grains. No signs of crystalline dust features have been found in R~Aqr so far. In ISO's long-wavelength range (LWS), the spectrum shows no prominent features. The ISO spectrograph had a wavelength-dependent field of view of $\approx 25\arcsec \times 25\arcsec$, thus not covering the outer regions observed by Herschel. To derive the Herschel-PACS fluxes of the central star, aperture photometry with a radius of $10\arcsec$ was applied. The resulting fluxes are $F_{\nu,70}=20.980$\,Jy and $F_{\nu,160}=2.569$\,Jy at 70\,$\mu$m and 160\,$\mu$m, respectively, and coincide well with the predictions of the combined photosphere and wind model (to be described below).
\begin{table}[t]
      \caption[]{Input parameters used for modelling the atmosphere and wind of R~Aqr.}
      \label{input_wind}
    \centering                                     
      \begin{tabular}{lc}          
\hline
\hline      
luminosity $L_\star$ [$L_{\sun}$] & 4780 \bigstrut[t] \\
effective temperature $T_{\rm eff}$ [K] & 2800 \\
mass $M_\star$ [$M_{\sun}$] & 1 \\
surface gravity log($g_\star$) [cm\,s$^{-2}$] & --0.5 \\
metallicity $Z$ [$Z_{\sun}$] & 1 \\
C/O ratio & 0.55 \\
mass-loss rate $\dot{M}$ [$M_{\sun}$\,yr$^{-1}$] & $7 \times 10^{-7}$ \bigstrut[b] \\
\hline
\end{tabular}
\end{table}

Three PACS fluxes are plotted in Fig.~\ref{R_Aqr_SED}. The first (highest) one corresponds to the overall IR emission (aperture photometry on a circle of radius $50\arcsec$ with resulting fluxes of $F_{\nu,70}=28.220$\,Jy and $F_{\nu,160}=4.320$\,Jy at 70\,$\mu$m and 160\,$\mu$m, respectively). The intermediate flux value corresponds to the above described measurement of the central source, while the third (weakest) one corresponds to the flux of the eastern arm. An elliptical aperture with a size of $40\arcsec \times 20\arcsec$ tilted by $45\degr$ was chosen to fully cover the arm. The resulting fluxes for the arm are $F_{\nu,70}=0.481$\,Jy and $F_{\nu,160}=0.288$\,Jy at 70 and 160\,$\mu$m, respectively. As discussed in more detail in Sect.~\ref{Sect:RAqrCSE}, this flux measurement in the CSE of R~Aqr may be interpreted as astronomical silicate dust emission.
Adopting absorption coefficients $Q_{\rm abs,70} = 3\times10^{-3}$ and $Q_{\rm abs,160} = 5.9\times10^{-4}$ at 70\,$\mu$m and 160\,$\mu$m, respectively \citep{Draine1985}, the 70\,$\mu$m and 160\,$\mu$m fluxes observed at the eastern tip of the eastern arm then correspond to a temperature of 37\,K \citep[see][for details]{Jorissen2011}. Because there is a temperature gradient in the arm (see Sect.~\ref{Sect:RAqrCSE}), the average temperature of the arm is $\approx 32$\,K. The corresponding  dust-emission spectrum -- computed via $\nu \pi B_{\nu}(T$\,=\,32${\rm K})Q_{\rm abs}(\nu) f$, where $f$ is a scaling factor -- is also shown in Fig.~\ref{R_Aqr_SED}.

To characterize R\,Aqr, we modelled the atmosphere and the inner wind region using methods of proven reliability. The model parameters, listed in Table~\ref{input_wind}, were chosen according to values compiled from the literature where available. With an angular diameter of $\theta_*=0\farcs015$ \citep{vanBelle1997}, a distance of $D=214$\,pc \citep{Kamohara2010}, and assuming the star to have a mass of 1\,$M_{\odot}$, one can estimate log($g_\star$). Together with the effective temperature adopted from \citet{Contini2003}, the luminosity of R~Aqr can be derived. Metallicity and C/O ratio were assumed to be solar-like following the abundance pattern derived by \citet{Caffau2008,Caffau2009}. For the stellar atmosphere, we computed a COMARCS model atmosphere following  \citet{Aringer2009}. This was then complemented with a model for the stationary outflow containing amorphous silicate dust particles, calculated as described in detail by \citet{Ferrarotti2006},  including recent improvements of the wind code (Gail et al., in prep.). 
Based on this combined model, we carried out detailed radiative transfer calculations similar to previous works \citep{Aringer2009,Nowotny2011}.

This led to a synthetic spectrum that simulates an unresolved object with spectral contributions from the photosphere and the inner-wind region (but not from any extended circumstellar material farther than the eastern arm of R~Aqr), just as can be assumed for the observational data in Fig.~\ref{R_Aqr_SED}. The resulting model spectra were normalised to the flux of the ISO-SWS spectrum at 3.8\,$\mu$m for a direct comparison. Overplotted in the figure is the best-fitting model after iterating the mass-loss rate. While the COMARCS model is able to reproduce the SED of R~Aqr in the visual and NIR reasonably well, an additional emission due to the circumstellar dust is clearly needed for a proper description beyond $\approx4$\,$\mu$m. Figure\,\ref{R_Aqr_SED} illustrates that our wind model with a final mass-loss rate of 7$\times$10$^{-7}$\,$M_{\odot}$\,yr$^{-1}$ is also able to reproduce the emission features due to amorphous silicates at 9.7 and 18\,$\mu$m quite well. This successful model will be used in Sect.~\ref{Sect:RAqrCSE} to compare the dust temperatures resulting from the wind model to the corresponding ones estimated from PACS fluxes.

\subsubsection{Infrared arms in R~Aqr: Jets, expanding nova, or ISM interaction?}
\label{Sect:RAqrCSE}

R~Aqr was observed on the 16th of May 2011 with the PACS photometer of Herschel at 70\,$\mu$m and 160\,$\mu$m. Figure~\ref{R_Aqr_deconv} shows the first successful attempts to image the far-IR emission around R~Aqr. Both images reveal the same features. Since the 70\,$\mu$m image offers a better resolution, however, it will be the base for the following discussion (unless otherwise noted).

The far-IR circumstellar environment is basically characterised by two arms, one approximately in north-east direction and the other in south-west direction. In the 160\,$\mu$m band, the eastern arm suffers from the lower resolution and is only visible as a bright clump, whereas the western arm sustains its shape. 
\begin{figure}[t]
\centering
   \includegraphics[width=9cm]{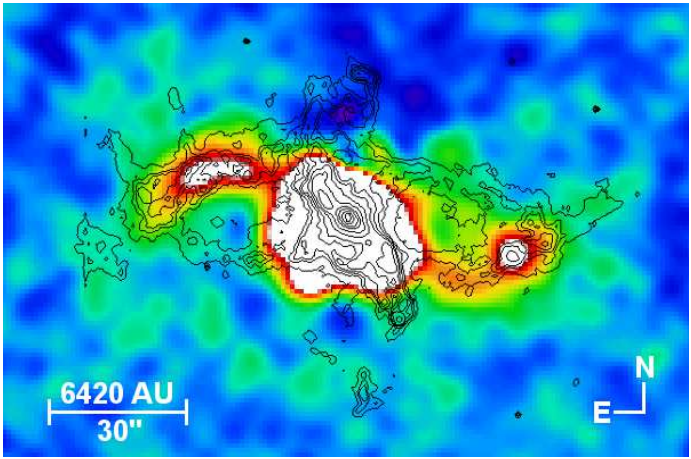} \\
     \caption{Deconvolved 70\,$\mu$m PACS image of R~Aqr overplotted with contours from an ESO Multi-Mode Instrument (EMMI) 2007 observation at 653\,nm [\ion{N}{ii}]. The contours clearly show a ring structure while the far-IR image renders only a part of this structure.}
\label{R_Aqr_contours}
\end{figure}

The arms of R~Aqr are exactly opposite at PA = 61$\degr$ and 241$\degr$ and seem to be connected with the binary system. However, owing to the saturation of the central source, a look farther inside a region closer than $\approx 18\arcsec$ from the star is not possible. The overall size of the IR emission structure around R Aqr is about $95\arcsec \times 35\arcsec (\approx 20330 \times 7490$\,AU). While both arms are aligned nearly symmetrically with respect to the central star, the light distribution is not. The western arm is brightest at its end, in the form of a light spot, whereas the eastern arm is clearly dominated by emission from the curved region. We now discuss three possible origins for the arms observed in the infrared.

\subsubsection*{Bipolar jet?}

The first hypothesis naturally arises from the morphology of the IR emission and assumes that the arms trace the bipolar jet from the symbiotic system (Sect.~\ref{R_Aqr_background}). This hypothesis is not well supported because of the discrepancy between the position angles of the IR arms (61$\degr$ and 241$\degr$) and of the optical-UV-X jet \citep[46$\degr$ and 211$\degr$;][]{Kellogg2007}. Observations in the UV by \citet{Meier1995} indicate that the direction of the bipolar wind is not stable due to the precession of the accretion disk. Although such a precession effect could perhaps account for the above discrepancy, we do not engage in discussing the possible association of the infrared arms with the jet, since the hypothesis cannot be validated with our observations.

\subsubsection*{Nova nebula?}

The second hypothesis about the nature of the infrared arms thus stems from their good agreement with the structures seen for instance in [\ion{N}{ii}] 653\,nm, as shown in Fig.~\ref{R_Aqr_contours}, which correspond to the equatorial plane of the outer hourglass structure first outlined by \citet{Solf1985} (see Sect.~\ref{R_Aqr_background}).\footnote{The jets at PA = 46$\degr$ and 211$\degr$ \citep{Kellogg2007} are clearly visible in the contour plot of Fig.~\ref{R_Aqr_contours} and do not match the far-IR emission. In the optical, the jets extend up to 20$\arcsec$ north and 30$\arcsec$ south from the star, well beyond the PACS far-IR emission, and have no IR counterpart, which is also true for the detached clump seen in the optical approximately 40$\arcsec$ south of the star and for the (non-detached) clump 37$\arcsec$ north.}

In the eastern arm, the optical emission is stronger on the northern half, which is moving away from the observer \citep{Hollis1999}. There it fits its far-IR counterpart well. In the southern part of the eastern arm, far-IR emission is missing, but the optical emission is also much weaker. Similar conclusions hold for the western arm: this time, it is the southern part of the ring that matches the IR emission (even the IR spot at the arm's end has a well-defined  optical counterpart), whereas the northern part of the ring is missing IR emission. This close match between optical and far-IR features raises the question of the origin of the far-IR emission. There are two possibilities: either scattering of light from the central source by the gas or dust, or else continuum dust emission. In the first case, the good match between the optical and far-IR images is a natural consequence of the model. The high degree of polarization observed from the nebula around R~Aqr \citep{Yudin2002} is in favour of this model. The second hypothesis requires the dust to be strongly coupled with the gas: it must either have formed in the nova ejecta traced by the gas seen in the optical image, or it must have been piled up by the ejecta in the Mira wind pre-existing to the explosion.
\begin{figure}[t!]
\centering
   \includegraphics[width=9cm]{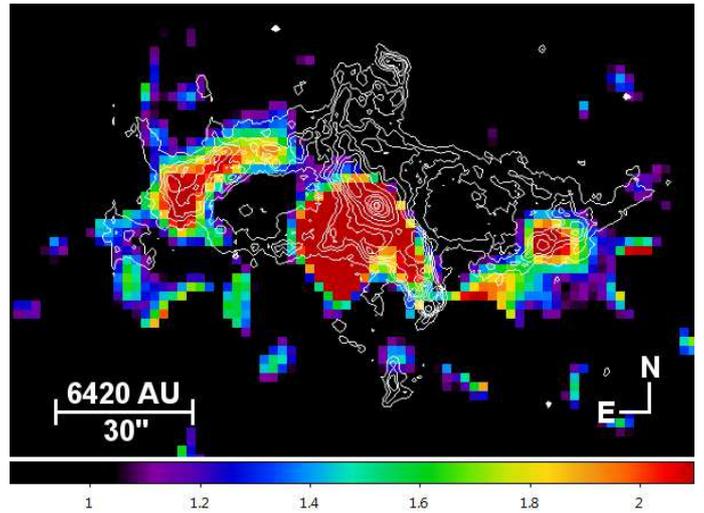} \\
     \caption{Map from the ratio of flux values between the deconvolved 70\,$\mu$m and 160\,$\mu$m PACS images of R~Aqr overplotted with contours from an EMMI observation at 653\,nm [\ion{N}{ii}]. The colour code represents the flux ratio between 70 and 160\,$\mu$m. Both arms show gradients in the flux ratio, which is a signature of continuum dust emission. Furthermore, the highest flux ratio occurs at the ends of both arms.}
\label{R_Aqr_temp}
\end{figure}

To distinguish between the two models -- light scattering or continuum dust emission -- a map of the 70\,$\mu$m to 160\,$\mu$m flux ratio was derived (Fig.~\ref{R_Aqr_temp}). Fluxes below 4$\sigma$ of the background noise in either band were set equal to  zero  to prevent artefacts originating from background fluctuations. The part of the western arm closest to the star suffers from artefacts in the 160\,$\mu$m band, which are not entirely eliminated by the 4$\sigma$ clipping. The image clearly shows gradients in both arms, which are not expected in the scattered-light scenario, since the flux ratio is a function of the scattering cross section only (provided that the scattering centres are the same throughout the arms) and is thus expected to be constant throughout the arms. 
 The flux ratio image of Fig.~\ref{R_Aqr_temp} thus seems in favour of dust emission rather than light scattering.  
Even more surprising, the map reveals that the highest ratio occurs at the end of both arms, which are situated opposite each other much more accurately than in Fig.~\ref{R_Aqr_contours}. At the end of the arms, the flux ratio is $\approx2.5$ while at the northernmost part of the eastern arm, the ratio drops to $\approx1.1$. To constrain the temperature from these flux ratios, it is necessary to specify the dust species in the arms.
 Assuming the dust to be constituted only of silicates with absorption coefficients $Q_{\rm abs,70} = 3\times10^{-3}$ and $Q_{\rm abs,160} = 5.9\times10^{-4}$ at 70\,$\mu$m and 160\,$\mu$m, respectively \citep{Draine1985}, the flux ratios observed in the eastern arm correspond to temperatures of 37\,K and 29\,K \citep[see][for details]{Jorissen2011}. Because astronomical silicate has a very low opacity according to \citet{Draine1985}, these values represent the lowest possible ones.
 
The temperature of the eastern arm tip may now be compared with the predictions from the modelling at the same distance, since the model for photosphere and dusty outflow has been shown to describe the resulting SED of central source and inner wind region quite satisfactorily (Fig.~\ref{R_Aqr_SED}). The wind model yields a silicate dust temperature of 49\,K at a distance of 9200\,AU. These values compare reasonably well to the estimates based on the observational data (see above), and the successful application of the wind model to predict the dust temperature implies moreover that the dust is heated by the radiation from the central star rather than by some form of shock.

\begin{figure}[t!]
\centering
   \includegraphics[width=9cm]{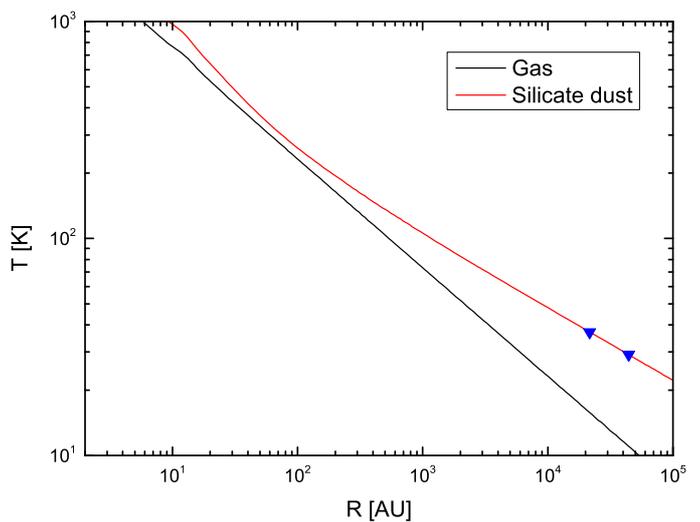} \\
     \caption{Temperature gradient delivered by the stationary wind model (see Sect.~\ref{R_Aqr_spectra}). The dust temperatures of the eastern arm derived from the PACS band photometry (37\,K and 29\,K; blue triangles) correspond to distances of 21\,500\,AU and 43\,700\,AU in the wind model.}
\label{R_Aqr_model_temp}
\end{figure}

The temperature \textit{gradient} predicted by the wind model (Fig.~\ref{R_Aqr_model_temp}) may now be used to de-project the geometry of the eastern arm. Since the (logarithmic) gradient appears to stay constant for distances larger than 500\,AU from the central star (i.e., well encompassing the location of the eastern arm), it is more robust than the temperature prediction itself (the dust temperature of 37\,K observed at a distance of 9200\,AU differs somewhat from the 49\,K value predicted by the model). The observed temperature ratio of 1.27 ($=37/29$) along the eastern arm is thus equivalent to a (de-projected) distance variation by a factor of 2.03 [$=d(29\,{\rm K})/d(37\,{\rm K})$]. Since the tip of the arm is its warmest part, it may be assumed that it is located closest to the star, by the factor 2.03 derived above, as compared to the coolest, northernmost point on the eastern arm. On the image, however, the corresponding (projected) distances are $d_{\rm proj} (29\,{\rm K}) = 4300$\,AU, and $d_{\rm proj} (37\,{\rm K}) = 9200$\,AU, or a ratio of 0.467. This is in contrast to the de-projected one, thus suggesting that the arm is strongly inclined onto the plane of the sky. To fix the ideas, we assume that the warmest tip of the eastern arm [at PA$ = 90\degr$, corresponding to $d_{\rm proj} (37\,({\rm K})$] lies in the plane of the sky (basically because the western and eastern tips are almost exactly opposite). With this assumption, we obtain $d_{\rm proj} (37\,{\rm K}) = d(37\,{\rm K})$ and  $d_{\rm proj} (29\,{\rm K}) / d (29\,{\rm K}) = \cos i = 0.467 / 2.03 = 0.23$, or $i = 77\degr$. Incidentally, this value is close to the $72\degr$ inclination of the orbital plane found by \citet{Solf1985}. Therefore, the picture emerging from our analysis for the arms is that of a non-circular structure located in the orbital plane. 

\subsubsection*{Interaction with the ISM?}

A third possible interpretation of the arm feature relates to its orientation with respect to the space motion of the star. Adopting the proper motion from the SiO maser measurements \citep{Kamohara2010} and removing the solar motion $(U,V,W)_{\sun} = (11.10,12.24,7.25)$~km\,s$^{-1}$ obtained by \citet{Schonrich2010}, we found that the star moves with $v_{\rm LSR} = 39.7$\,km\,s$^{-1}$ along PA = 145.1$\degr$ with an inclination $i = -40.2\degr$ with respect to the plane of the sky. In Fig.~\ref{R_Aqr_deconv}, the space motion is indicated by an arrow and its length gives the position of the star 2000 years in the future. The space motion thus appears to be almost perpendicular to the arms (PA = 61$\degr$). It is worth noting that a similar orthogonality has been found between the jet and the space motion of X~Her \citep{Jorissen2011}. However, unlike the situation prevailing for X~Her, the arm structure does not seem to be linked to the space motion, since ram pressure would bend the arms in the direction opposing the star motion, unlike what is observed for the eastern arm. Because of this, it is very unlikely that the ISM shapes the arm structure.
\\
\\
As a conclusion to the discussion of these three possible models, the arms seem to be the IR counterpart of the expanding nebula associated with the outer hourglass observed in the optical and the emission is caused by radiation-heated dust grains. However, the temperature gradient within the eastern arm, along with the exactly opposing locations of the highest temperatures, seems to contradict the circular geometry of the expanding ring proposed by \citet{Solf1985}, since this model would imply if not a uniform temperature distribution, at least a more arbitrary one than that observed. Although \citet{Solf1985} clearly favoured the interpretation of an inclined ring ejecta, they also mentioned (their Sect.~4.3) the possibility that the nebula could be elliptical, based on kinematic arguments. This interpretation is supported by the temperature map of the Herschel observations.

\subsection{W~Aql}
\label{W_Aql}

\subsubsection{Background}

W~Aql is a Mira variable star of spectral type S6.6 with a period of 490 days \citep{Feast2000} at a distance of 340\,pc \citep{Guandalini2008}. 
W~Aql was found to be a binary star by \citet{Herbig1965}, who noticed that the spectrum at minimum light acquired spectral characteristics of an F5 or F8 star. \citet{Ramstedt2011} recently presented an archive HST 435\,nm image from 2004 that clearly shows the companion south-west of W~Aql at a distance of $0\farcs47$. This angular separation corresponds to a projected distance of 160\,AU at 340\,pc.

W~Aql shows strong emission lines of SiO and HCN with a parabolic shape indicating that the emission is optically thick \citep{Bieging1998}. In addition, \citet{Ramstedt2009} found weak wings extending beyond the parabolic profile. They could be caused by recent mass-loss-rate modulations or asymmetric outflows.

The mid-IR spectrum of W~Aql is also dominated by emission from amorphous silicate grains, with an additional contribution from amorphous alumina dust. Both emissions result in a peak position (for the blended silicate plus alumina bands) of about 10.5\,$\mu$m \citep{Yang2007}.

The circumstellar environment was studied at 11.15\,$\mu$m  by \citet{Danchi1994} and \citet{Tatebe2006}. The former authors detected dust shells at 8 and 19 stellar radii while the latter found the dust to be unevenly distributed, with a clear excess eastward. 

\begin{figure}[t]
\centering
   \includegraphics[width=9cm]{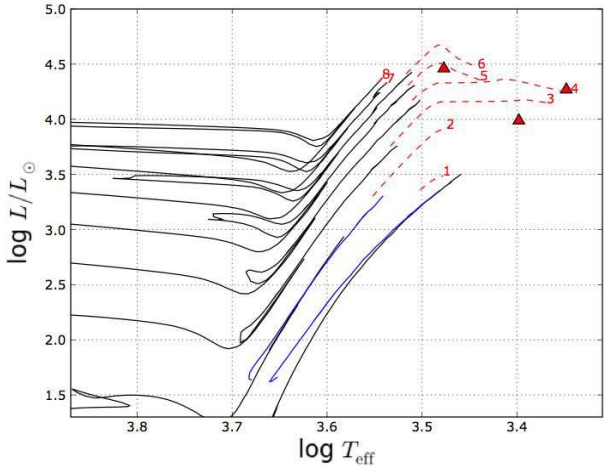}
     \caption{Evolutionary tracks from the Padova set \citep{Bertelli2008,Bertelli2009} for $Y$=0.26 and $Z$=0.017, with masses between 1 and 8\,$M_{\sun}$, as indicated (in red) at the end of each track. The asymptotic giant branches are depicted with dashed (red) lines. The location of W~Aql as derived  from the interferometric radii measured by \citet{Danchi1994} and \citet{vanBelle1997} are depicted  by (red) triangles.}
\label{Fig:tracks_WAql}
\end{figure} 

\subsubsection{Nature of the companion}
\label{W_Aql_companion}

The suspicion of binarity based on the composite spectrum at minimum light discovered by \citet{Herbig1965} was later confirmed by the HST images of \citet{Ramstedt2011}. The log file of the available HST images is given in Table~\ref{Tab:HST}. Two sets of images separated by 11\,yr are available and can be used to check for a possible relative motion of the two components.  None has been detected, though, within the error bar set by the pixel size, which is $0\farcs04$ for the WFPC, about two times larger than that of the ACS image. At a distance of  340\,pc, this pixel size  corresponds to 13.6\,AU. Assuming a circular orbit with a radius of 160\,AU (as derived from the angular separation of $0\farcs47$), the pixel size represents about 1.3\,\% of the orbital circumference. Over a 11\,yr timespan, a motion would be detectable with the HST if the orbital period were shorter than about 800\,yr.
\begin{figure*}[t!]
\centering
   \includegraphics[width=9cm]{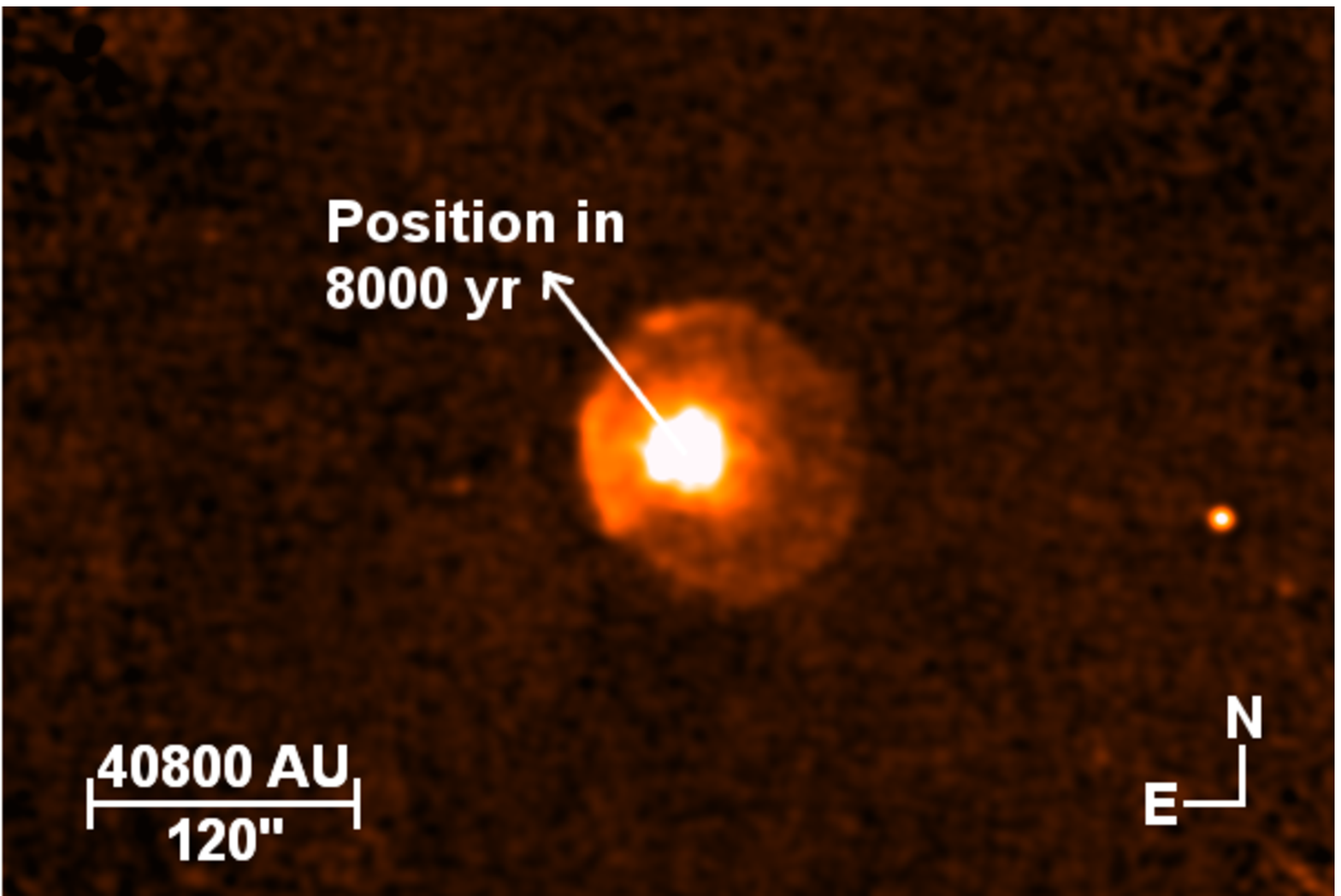}
   \includegraphics[width=9cm]{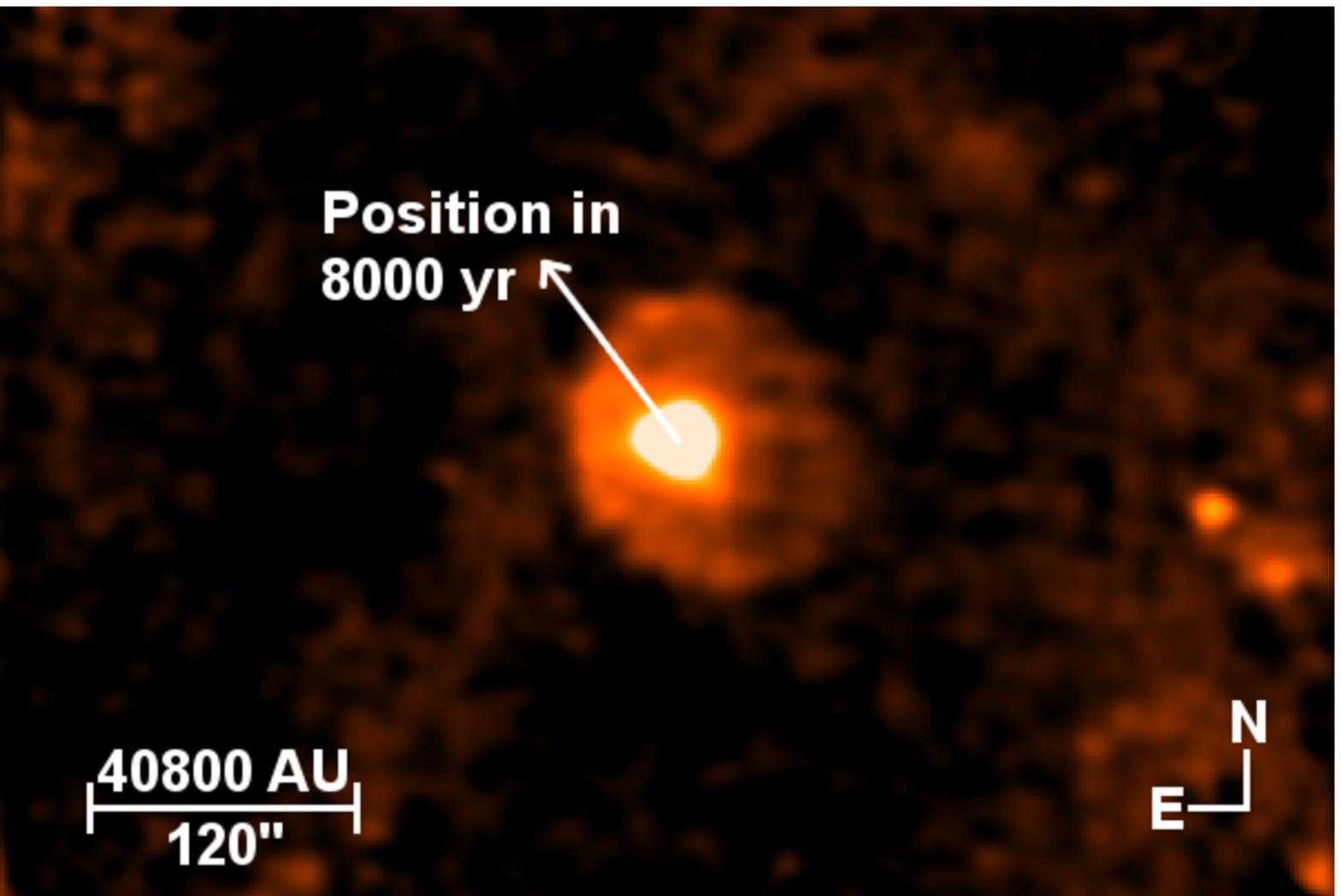}
     \caption{Deconvolved PACS images of W~Aql at 70\,$\mu$m (left) and at 160\,$\mu$m (right). The arrows indicate the direction of space motion and the position of the star 8000\,yr in the future.}
\label{W_Aql_deconv}
\end{figure*} 

A coincidental accord with a background source can also be excluded. The distance covered over 11\,yr at a distance of 340\,pc by an AGB star moving (rather slowly) with $v_{\rm LSR} = 20$\,km\,s$^{-1}$ is  more than a factor of 3 larger than the pixel size of the WFPC image ($0\farcs04$ compared to $0\farcs13$ covered by the star). 
\begin{table}\footnotesize
\caption{\label{Tab:HST}
Images of W Aql retrieved from the HST archives, with column labelled ID providing the HST program number. Column $\Delta m$ provides the magnitude difference ($m_{\rm W Aql} - m_{\rm comp}$) as estimated from aperture photometry. Columns labelled $V$ and $V_{\rm comp}$ provide the visual magnitudes of W~Aql as derived from the AAVSO (for 1993) or ASAS (for 2004) databases \citep{Pojmanski2002}, and of its companion, respectively. The last column gives the measured separation $\delta$.}
\begin{tabular}{rrrrrrrrrrrr}
\hline
\hline
Instr. & Date & ID & $\Delta m$ &  $V$ & $V_{\rm comp}$ & $\delta$ \bigstrut[t] \\
Filter              &        &                 &       &  & & ($\arcsec$) \\
\hline
WFPC & 1993-05-25 & 3603 & +0.30 & & & \bigstrut[t] \\
F487N  & & & & & & \\
\\
WFPC & 1993-05-25 & 3603 & -0.99 & 13.8 & 14.8 & 0.47 \\
F502N & & & & & & \\
\\
ACS & 2004-10-12 & 10185 & -3.53 & & & \\
F435W  & & & & & & \\
\\
ACS & 2004-10-12 &  10185 & -3.82 & 11.1 & 14.9 & 0.47\\
F606W & & & & & & \\
\hline
\end{tabular}
\end{table}
To derive the nature of the companion, aperture photometry has been performed on the two continuum images (obtained with filters F502N and F606W), and the results are listed in Table~\ref{Tab:HST} in the form of a magnitude difference with respect to W~Aql. 

The magnitude of the Mira variable at the time of the HST observations was obtained from the AAVSO database (for the 1993 observation) or from the ASAS database \citep{Pojmanski2002} (for the 2004 observation). 
It is remarkable that both epochs yield a consistent value for the visual magnitude of the companion (about 14.8), and that this magnitude is well below the magnitude of  W~Aql  at minimum light ($V_{\rm W Aql, min} \approx 14.0$). This explains the absence of a flat minimum for  the light curve, which would be caused by  the companion becoming  brighter than the  variable star  around minimum light.

Therefore, adopting an apparent magnitude of  14.8 for the companion and a distance of 340\,pc, the resulting absolute visual magnitude is $M_V = 7.1$, corresponding to a K4V main-sequence star with a mass of $\approx 0.7$\,$M_{\sun}$. This result contrasts with the F5 or F8 spectral type inferred by \citet{Herbig1965} from the composite spectrum observed at minimum light;  but the spectrum quality was quite poor.

We now estimate the mass of W~Aql to evaluate the orbital period. \citet{Danchi1994} and \citet{vanBelle1997} determined the angular diameter $\theta$ and effective temperature of W~Aql. Van Belle et al. (1997) observed the star at maximum light and found $\theta = 11.58$\,mas and $T_{\rm eff} = 2230$\,K, which yields $L = 18\,590$\,$L_{\sun}$ at a distance of 340~pc. \citet{Danchi1994} observed the star at maximum and minimum light and found $\theta_{\max} = 8.6$\,mas at $T_{\rm eff,max} = 3000$\,K and $\theta_{\rm min} = 7.2$\,mas at $T_{\rm eff,min} = 2500$\,K, corresponding to $L_{\rm max} = 28\,885$\,$L_{\sun}$ and $L_{\rm min} = 9740$\,$L_{\sun}$. These values are widely spread and lead to a stellar mass in the range 2 to 5\,$M_{\sun}$ using the Padova stellar evolutionary tracks for a solar composition (Fig.~\ref{Fig:tracks_WAql}). This yields a total system mass in the range 2.7 -- 5.7\,$M_{\sun}$. Adopting a circular orbit of radius 160\,AU ($0\farcs47$ at 340~pc and $i = 0$), a minimum orbital period  in the range 850 -- 1230\,yr is found from Kepler's third law. This estimate agrees with that based on the absence of any relative motion on the HST image and has to be seen as the lowest bound, since the de-projected radius of the orbit can be much larger than 160\,AU.

\subsubsection{Space motion} 
\label{W_Aql_space_motion}

\begin{figure}[t]
\centering
   \includegraphics[width=9cm]{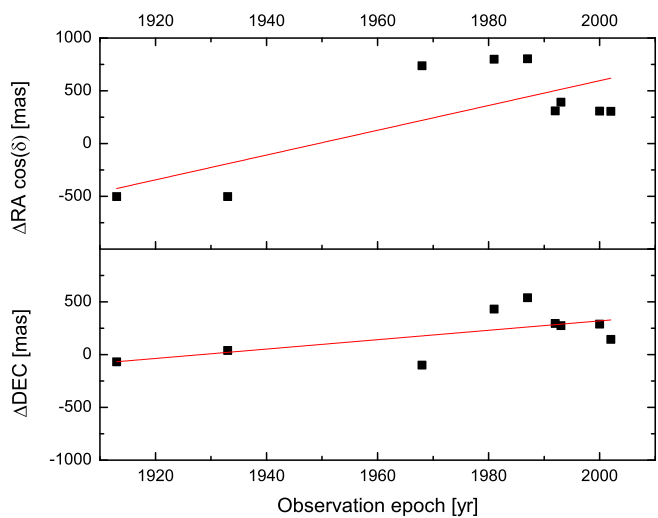}
     \caption{Positional data of W~Aql. The positions are taken from 9 catalogues (see Table \ref{proper_motion_table} and text) and cover the years 1913 to 2002. The (red) lines indicate the linear fit to the data points. The scatter is of the order of the angular separation of the binary system on the sky that is expected for a variability-induced movement. The positions of W~Aql in the last century show a trend for a movement along PA $= 38\degr$ with a space velocity $v_{\rm LSR} = 21.6 \pm 4.1$\,km\,s$^{-1}$.}
\label{W_Aql_positions}
\end{figure}

Unfortunately, the space motion of W~Aql is not clearly defined, as shown in Table~\ref{proper_motion_table}, which collects proper-motion values from various catalogues. Unlike the situation prevailing for the proper motions of R~Aqr, also listed in Table~\ref{proper_motion_table}, the proper motion of W~Aql varies significantly from one catalogue to the other, with no trend with the mean epoch spanning almost a century. 

This behaviour very likely results from the combination of the Mira pulsation and the presence of a companion. 
During the variability cycle of the Mira, the photocenter of the binary system moves along the path that joins the two stars (separated by 0.47$\arcsec$), according to the respective contribution of the two components to the total light
(according to AAVSO, the Mira variable varies from $V = 7$ to 14, while the companion has $V \sim 14.8$ as shown above). The proper motion, of the order of a few tens of milliarcseconds per year, can therefore not be reliably derived, since the individual position measurements on which it relies are varying over a range of about 400 milliarcseconds, with a periodicity of 490~d. This behaviour is denoted as ``variability-induced mover" (VIM) in the context of the Hipparcos catalogue \citep[see also][]{Pourbaix2003}.

Nevertheless, on a century-long time span, one may hope that the long-term proper motion will dominate over this short-term variability-induced motion. For this reason one can try to derive the proper motion from the positional data of the seven catalogues in Table \ref{proper_motion_table} (see references within) for which an observation epoch is stated. The SAO Star Catalog has to be excluded because W~Aql's position is far off from that of the other catalogues and seems erroneous. Instead, the positions from the HST Guide Star Catalog \citep[obs. epoch 1987;][]{Lasker1996}, the Tycho-2 Catalog \citep[obs. epoch 1991;][]{Hog2000} and the USNO-B Catalog \citep[obs. epoch 2000;][]{Monet2003} are added. They are not listed in Table \ref{proper_motion_table} because they do not feature proper motion values for W~Aql. The result are nine data points that cover nearly one century (1913--2002). Figure~\ref{W_Aql_positions} shows the right ascension and declination values with respect to the observation epoch. The diagrams clearly show a trend for the stellar motion of W~Aql. Linear curve fitting was performed to obtain the proper motion from the slopes of the curves. The resulting proper motion values are $\mu_{* \alpha} = 11.8 \pm 4.1$\,mas\,yr$^{-1}$ and $\mu_{\delta} = 4.5 \pm 2.1$\,mas\,yr$^{-1}$ for right ascension and declination, respectively. These values are different from all stated in Table \ref{proper_motion_table} but reflect observations of the stellar position over a timespan of nearly one century, essentially longer than the observations for every catalogue in the table. In addition, the direction of motion is largely supported by the IR observations of the CSE, which are discussed in Section \ref{W_Aql_PACS}. 
\begin{figure*}[t!]
\centering
   \includegraphics[width=9cm]{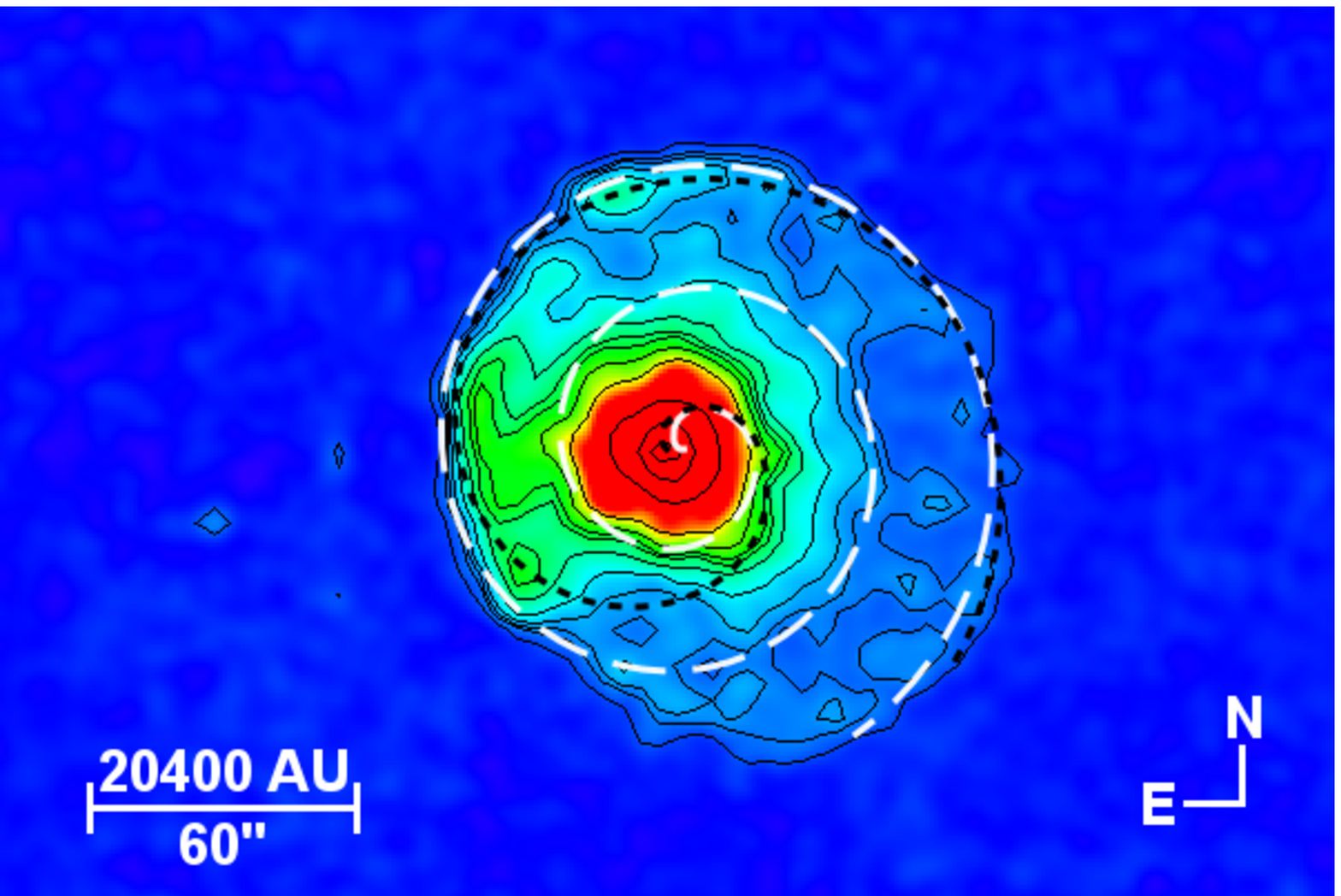}
   \includegraphics[width=9cm]{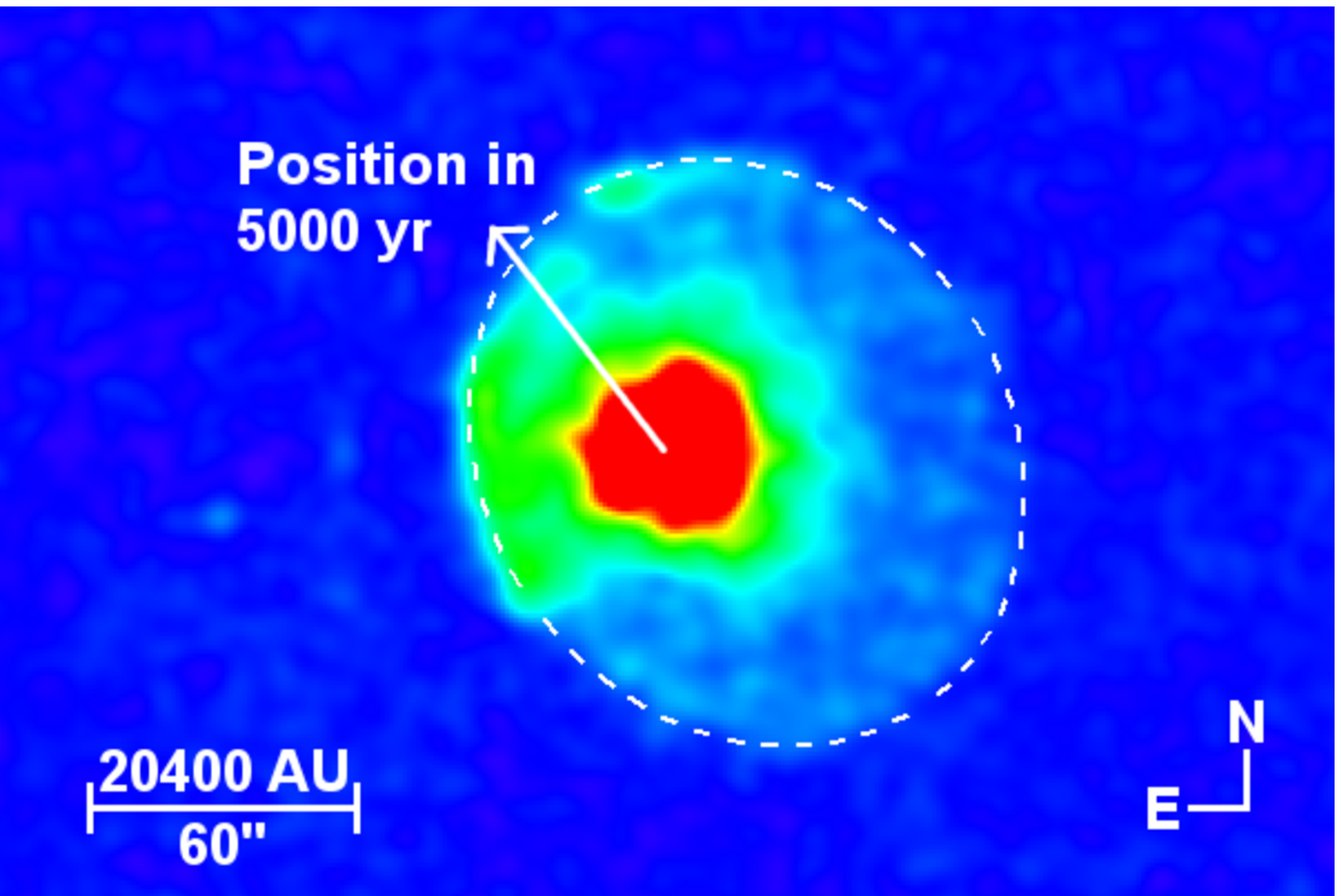}
     \caption{Left panel: Contour plot of the deconvolved 70\,$\mu$m PACS image of W~Aql. Two Archimedean spirals matching the outline of the IR emission are recognizable. The (black) dotted spiral has one winding to cover the emission with a separation distance $\theta = 52\arcsec$ while the (white) dashed spiral has two windings with a separation distance $\theta = 26\arcsec$. Right panel: the same image overplotted with an ellipse fitted to the IR emission. The axes lengths of the ellipse are 69$\arcsec$ and 59$\arcsec$, respectively, with the major axis oriented along PA$=35\degr$.}
\label{W_Aql_contours}
\end{figure*}

Although the scatter around the linear fit seems large at first sight, it is $\approx 2 \times 10^{-4}$ degrees or $\approx 0\farcs7$, which is about the separation of the binary system ($0\farcs5$). This is exactly what is expected for a variability-induced movement, whose amplitude must be of the order of the angular separation on the sky. 

Adopting a distance of 340\,pc \citep{Guandalini2008}, a radial velocity $V_{\rm r} = -20.80 \pm 3.10$\,km\,s$^{-1}$ \citep{Kharchenko2007} and the solar motion $(U,V,W)_{\sun} = (11.10,12.24,7.25)$\,km\,s$^{-1}$ \citep{Schonrich2010}, the resulting proper motion values are $\mu_{* \alpha} = 7.9 \pm 4.1$\,mas\,yr$^{-1}$ and $\mu_{\delta} = 10.0 \pm 2.1$\,mas\,yr$^{-1}$ for right ascension and declination, respectively. The space velocity of W~Aql is $v_{\rm LSR} = 21.6 \pm 4.1$\,km\,s$^{-1}$ at PA$= 38\degr$ with an inclination $i = -17\degr$ with respect to the plane of the sky. However, due to the large uncertainties of the positions, the results thus obtained may include large errors.

\subsubsection{Binary vs. ISM interaction}
\label{W_Aql_PACS}

W~Aql was observed with PACS on October 2011. Figure~\ref{W_Aql_deconv} presents the 70\,$\mu$m and 160\,$\mu$m images. Since they exhibit the same features, the discussion below relies on the 70\,$\mu$m image.\\
In the IR, the CSE is at first sight a typical \textit{fermata} case and was classified as such by \citet{Cox2012}. The elliptical envelope shape is to be expected from the interaction of the stellar wind with the oncoming ISM and the orientation of the envelope with its symmetry axis at PA $\approx 35\degr$ (with the star offset by 8$\arcsec$ though) nicely fits the space motion of the star along PA $= 38\degr$ (see right panel of Fig~\ref{W_Aql_contours}). 

A feature that does not directly fit with this interpretation though is the brighter region east of the star, at a distance of 42$\arcsec$. This feature suggests a bow shock but its location and the space motion/symmetry axis of the ellipse do not match: the bright region eastward extends from PA = $55\degr$ to $140\degr$ while the stellar motion and the CSE have their symmetry axis along PA $\approx 35\degr$. However, it is imaginable that a strong ISM flow blowing from the east modifies the CSE morphology and creates a fermata. Another possible scenario is an ISM flow from the south that pushes material to the east. A feature supporting this is the bending of the IR emission $\approx28\arcsec$ south of the star towards the star. Such ISM flows combined with the proper motion of the star were proposed for $\alpha$~Ori by \citet{Ueta2008}. To infer the direction of the ISM flow, it is necessary to know the three-dimensional orientation of the bow shock, which is virtually impossible since the inclination of the bow shock cone has almost no effect on the (observed) projected outline \citep{Maclow1991,Cox2012}. Thus, the hypothesis of an ISM flow that interacts with the CSE at the bright region could be adequate, but it is hardly verifiable.
\\
\\
It is hard to imagine, however, that an ISM flow that creates such a strong shocked interface does not modify the overall symmetry of the CSE, which is aligned with the (quite slow) space motion of the star. The binary nature of W~Aql made us check whether a companion-induced Archimedean spiral could fit the outline of the CSE, as seen around RW~LMi \citep{Mauron2006} and LL~Peg \citep{Dinh2009}. \citet{Mastrodemos1999} showed in their hydrodynamic simulations that for large binary separations, such spirals even appear as broken arcs or concentric shells when seen edge-on \citep{Mayer2011}. The fact that such structures are not restricted to face-on orbits increases the probability to observe them.

Given the orbital period $P \geq 1000$\,yr and the wind velocity $v_{\rm w} = 20$\,km\,s$^{-1}$ (Section~\ref{W_Aql_companion} and Table~\ref{binary_list}), an arm separation 
\begin{equation}
\rho = v_{\rm w} \times P \geq 12\farcs5
\end{equation}
is estimated. Within this range, we indeed find two solutions, as drawn on the left panel of Fig.~\ref{W_Aql_contours}, with arm separations of 26$\arcsec$ or 52$\arcsec$, the former being about twice the estimated (minimum) winding separation. On the western side, the shell around W~Aql exhibits three equidistant sharp ridges that are quite clearly delineated by the 26$\arcsec$ Archimedean spiral. That spiral is broken on its southern side, however. An orbital period of 2000\,yr is thus required to match the Archimedean spiral, which is certainly not impossible, given the uncertainties associated with the distance and the orbital inclination.  

In addition, as indicated by \citet{Raga2011}, the starting point of the spiral structure ought not to be at the centre of the binary, but is offset by a distance 
\begin{equation}
R_0 = \frac{r_{\rm orb} \times v_{\rm w}}{v_{\rm orb}},
\end{equation}
where $r_{\rm orb}$ is the orbital separation, and $v_{\rm orb}$ is the orbital velocity. Within this distance, the wind propagates shock-less. For the W~Aql binary, the estimated orbital separation is 160\,AU and the orbital velocity 2.4\,km\,s$^{-1}$ (assuming an orbital period of 2000\,yr). This gives a shock-free radius of 1330\,AU or 4$\arcsec$.

In summary, there are good arguments to support both hypotheses of a binary-induced spiral 
and of a wind-ISM interaction. Although the Archimedean spiral nicely  fits the CSE, it cannot explain the bright region in the east. Similarly, an ISM flow alone cannot explain the CSE shape that follows an Archimedean spiral. Therefore, it is suggested that an ISM flow, preferentially from east or south direction, interacts with a part of the spiral and forms the bright region. A similar combination of orbital spiral and ISM interaction 
was already reported around $o$~Cet \citep{Mayer2011}.


\section{Conclusions}

We have analysed the properties of the 18 targets among the 78 AGB stars and RSGs in the MESS sample that are suspected of being binaries. Based on the \citet{Cox2012} classification of the circumstellar morphology into five groups, the binary objects tend to be dominantly of \textit{irregular} type (no imprint of the wind-ISM interaction on the circumstellar morphology). In contrast, the \textit{ring}-type category contains only one binary star, which suggests that the interaction of the companion with the stellar wind decisively alters the spherical symmetry of the wind. However, despite the small number of binary stars in the sample, we showed that the null hypothesis of a similar binary fraction among the irregular shells and among the total sample can be rejected at a 98\.\% confidence level.

Among the 17 binary AGB stars of the sample, a detailed analysis of the PACS images at 70\,$\mu$m and 160\,$\mu$m was performed for the two objects R~Aqr and W~Aql. 

R~Aqr, a symbiotic star, shows two arms extending over a region $95\arcsec \times 35\arcsec$ wide. The arms are exactly opposite at PA = 61$\degr$ and 241$\degr$, but their position does not fit the X-ray jets \citep[PA $41\degr$ and $211\degr$;][]{Kellogg2007}. However, a contour plot drawn from an ESO Multi Mode Instrument (EMMI) optical image shows that the IR arms partly fit the ring nebula described by \citet{Solf1985}. A map of the flux ratios between 70\,$\mu$m and 160\,$\mu$m, which is equivalent to a temperature map, reveals a gradient in both arms with the highest temperature at their ends. This result dismisses a circular-ring structure, which would not show any temperature gradient in the arms, and favours an elliptically-shaped nebula. In addition, the SED of R~Aqr (constructed with observational data compiled from the literature) was fitted with the help of a combined model for photosphere and dusty outflow, the latter having a mass-loss rate of $\dot{M}$\,$\approx$\,7$\times$10$^{-7}$\,$M_{\sun}$\,yr$^{-1}$. A silicate dust temperature of 37-–49\,K for the tip of the eastern arm could be estimated from the PACS photometry as well as from the modelling of the stellar wind. The temperature gradient delivered by the wind model results in an inclination of the arms by $77\degr$, which agrees well with the inclination of the orbital plane \citep[$72\degr$;][]{Solf1985}. Therefore, the arms are part of a non-circular structure located in the orbital plane.

For the second object, W~Aql, \citet{Ramstedt2011} presented an HST 435\,nm-image of the companion,  which was previously detected spectroscopically as an F5 or F8 star \citep{Herbig1965}. The quality of Herbig's spectrum is quite poor, however. Aperture photometry on archive HST images at other wavelengths (502 and 606\,nm) and light-curve data reveal that the companion has an absolute magnitude $M_V = 7.1$, corresponding to a K4 main-sequence star with a mass of $\approx 0.7$\,$M_{\sun}$. With a mass for W~Aql between 2 and 5\,$M_{\sun}$ (as determined from HRD track matching on luminosity and angular diameter data by \citet{Danchi1994} and \citet{vanBelle1997}), one derives an orbital period of at least 1000\,yr assuming a separation of 160\,AU. The outline of W~Aql's far-IR emission can be reproduced by an Archimedean spiral that has either one or two windings, with an arm separation of 52$\arcsec$ or 26$\arcsec$, respectively. A separation of 26$\arcsec$ would imply that the orbital period is about twice as large as the inferred 1000\,yr. However, this value for the period assumed an orbit seen face-on, but the binary separation (and hence the orbital period) may be much larger if the orbital plane is inclined with respect to the plane of the sky. The two-arm Archimedean spiral thus appears as a viable solution. But the CSE also shows some features, especially a bright region east of the star that can only be explained by the interaction with the ISM. Although the symmetry axis of the IR emission is along the space motion at PA$=36\degr$ and differs from the bright region at PA $55$--$140\degr$, an ISM flow from the east or south-east directions is a valid explanation for this feature. Thus the wind of W~Aql is most probably shaped both by the companion, which produces the spiral, and by the ISM, which is responsible for the spiral brightening upstream of the ISM flow. In a forthcoming paper, the binaries $\theta$~Aps and $\pi^1$~Gru will be analysed along similar guidelines.

\begin{acknowledgements}
We acknowledge the variable-star observations from the AAVSO International Database contributed by observers worldwide and used in this research. This work was supported in part by the
Belgian Federal Science Policy Office via the PRODEX Programme of ESA. AM, FK and MM acknowledge funding by the Austrian Science Fund FWF under project number P23586, RO under project number I163-N16 and WN under project numbers P21988-N16 and P23006. This work used data from the ESO archive (Program 079.D-0138) and from the HST archive (Programs ID 3603 and 10185). 
\end{acknowledgements}

\bibliographystyle{aa}
\bibliography{biblio}
\end{document}